\documentclass[12pt,preprint]{aastex}

\newcommand{\bdv}[1]{\mbox{\boldmath$#1$}}

\def\au{{\rm AU}}

\def\masyr{{\rm mas}\,{\rm yr}^{-1}}
\def\kpc{{\rm kpc}}
\def\mas{{\rm mas}}

\def\muas{\mu{\rm as}}

\def\max{{\rm max}}

\def\rel{{\rm rel}}

\def\e{{\rm E}}
\def\bpi{{\bdv\pi}}
\def\bmu{{\bdv\mu}}

\begin{document}
\title{OGLE-2016-BLG-1190Lb: First {\it Spitzer} Bulge Planet Lies Near the Planet/Brown-Dwarf Boundary}

\author{\textsc{Y.-H. Ryu$^{1}$, J. C. Yee$^{2}$, A. Udalski$^{3}$,
I.A.~Bond$^{4}$, Y.~Shvartzvald$^{5,^{\dag}}$, W.~Zang$^{6,7}$, R.
Figuera Jaimes$^{8,9}$, U.G. J{\o}rgensen$^{10}$, W. Zhu$^{11}$, C.
X. Huang$^{12,13,14}$, Y. K. Jung$^{2}$ \and M. D. Albrow$^{15}$,
S.-J. Chung$^{1,16}$, A. Gould$^{1,11,17}$, C. Han$^{18}$, K.-H.
Hwang$^{1}$, , I.-G. Shin$^{2}$, S.-M. Cha$^{1,19}$, D.-J.
Kim$^{1}$, H.-W. Kim$^{1}$, S.-L. Kim$^{1,16}$, C.-U. Lee$^{1,16}$,
D.-J. Lee$^{1}$ Y. Lee$^{1,19}$, B.-G. Park$^{1,16}$,
R. W. Pogge$^{11}$ \\
(KMTNet Collaboration)\\
S. Calchi Novati$^{20,21}$, S. Carey$^{22}$, C. B. Henderson$^{23}$,
C. Beichman$^{23}$,
B. S. Gaudi$^{11}$\\
({\it Spitzer} team)\\
P. Mr\'{o}z$^{3}$, R. Poleski$^{3,11}$, J. Skowron$^{3}$, M. K.
Szyma\'{n}ski$^{3}$, I. Soszy\'{n}ski$^{3}$, S. Koz{\l}owski$^{3}$,
P. Pietrukowicz$^{3}$, K. Ulaczyk$^{3}$,
M. Pawlak$^{3}$\\
(OGLE Collaboration)\\
F.~Abe$^{24}$, Y.~Asakura$^{24}$, R.~Barry$^{25}$,
D.P.~Bennett$^{25}$, A.~Bhattacharya$^{24}$, M.~Donachie$^{26}$,
P.~Evans$^{26}$, A.~Fukui$^{27}$, Y.~Hirao$^{28}$, Y.~Itow$^{24}$,
K.~Kawasaki$^{28}$, N.~Koshimoto$^{28}$, M.C.A.~Li$^{26}$,
C.H.~Ling$^{4}$, K.~Masuda$^{24}$, Y.~Matsubara$^{24}$,
S.~Miyazaki$^{28}$, Y.~Muraki$^{24}$, M.~Nagakane$^{28}$,
K.~Ohnishi$^{29}$, C.~Ranc$^{25}$, N.J.~Rattenbury$^{26}$,
To.~Saito$^{30}$, A.~Sharan$^{26}$, D.J.~Sullivan$^{31}$,
T.~Sumi$^{28}$, D.~Suzuki$^{25,32}$, P.J.~Tristram$^{33}$,
T.~Yamada$^{34}$,
T.~Yamada$^{28}$, A. Yonehara$^{34}$\\
(MOA Collaboration)\\
G.~Bryden$^{5}$, S.~B.~Howell$^{35}$,
S.~Jacklin$^{36}$\\
(UKIRT Microlensing Team)\\
M. T. Penny$^{11,^{\dag\dag}}$, S. Mao$^{6,37,38}$, Pascal
Fouqu\'e$^{39,40}$, T. Wang$^{6}$\\
(CFHT-K2C9 Microlensing Survey group)\\} R. A. Street$^{41}$, Y.
Tsapras$^{42}$, M. Hundertmark$^{42,10}$, E. Bachelet$^{41}$, M.
Dominik$^{8,^{\dag\dag\dag}}$, Z. Li$^{41}$, S. Cross$^{41}$, A.
Cassan$^{43}$, K. Horne$^{8}$, R. Schmidt$^{42}$, J.
Wambsganss$^{42}$, S. K. Ment$^{42}$, D. Maoz$^{44}$, C.
Snodgrass$^{45}$, I. A.
Steele$^{46}$\\
(RoboNet Team)\\
V. Bozza$^{21,47}$, M.J. Burgdorf$^{48}$, S. Ciceri$^{49}$, G.
D'Ago$^{50}$, D.F. Evans$^{51}$, T.C. Hinse$^{1}$, E. Kerins$^{38}$,
R. Kokotanekova$^{52,53}$, P. Longa$^{54}$, J. MacKenzie$^{10}$, A.
Popovas$^{10}$, M. Rabus$^{55}$, S. Rahvar$^{56}$, S.
Sajadian$^{57}$, J. Skottfelt$^{58}$ J. Southworth$^{51}$,
C. von Essen$^{59}$\\
(MiNDSTEp Team)\\}

\affil{$^{1}$Korea Astronomy and Space Science Institute, Daejon
34055, Korea}

\affil{$^{2}$Harvard-Smithsonian Center for Astrophysics, 60 Garden
St., Cambridge, MA 02138, USA}

\affil{$^{3}$Warsaw University Observatory, Al. Ujazdowskie 4,
00-478 Warszawa, Poland}

\affil{$^{4}$Institute of Natural and Mathematical Sciences, Massey
University, Auckland 0745, New Zealand}

\affil{$^{5}$Jet Propulsion Laboratory, California Institute of
Technology, 4800 Oak Grove Drive, Pasadena, CA 91109, USA}

\affil{$^{6}$Physics Department and Tsinghua Centre for
Astrophysics, Tsinghua University, Beijing 100084, China}

\affil{$^{7}$Department of Physics, Zhejiang University, Hangzhou,
310058, China}

\affil{$^{8}$Centre for Exoplanet Science, SUPA School of Physics \& Astronomy,
University of St Andrews, North Haugh, St Andrews KY16 9SS, UK}

\affil{$^{9}$European Southern Observatory, Karl-Schwarzschild-Str.
2, 85748 Garching bei M\"unchen, Germany}

\affil{$^{10}$Niels Bohr Institute $\&$ Centre for Star and Planet
Formation, University of Copenhagen, {\O}ster Voldgade 5, 1350 -
Copenhagen K, Denmark}

\affil{$^{11}$Department of Astronomy, Ohio State University, 140 W.
18th Ave., Columbus, OH 43210, USA}

\affil{$^{12}$Department of Physics and Kavli Institute for
Astrophysics and Space Research, Massachusetts Institute of
Technology, Cambridge, MA 02139, USA}

\affil{$^{13}$Dunlap Institute for Astronomy and Astrophysics,
University of Toronto, Toronto, ON M5S 3H4, Canada}

\affil{$^{14}$Centre of Planetary Science, University of Toronto,
Scarborough Campus Physical $\&$ Environmental Sciences, Toronto,
M1C 1A4, Canada}

\affil{$^{15}$University of Canterbury, Department of Physics and
Astronomy, Private Bag 4800, Christchurch 8020, New Zealand}

\affil{$^{16}$Astronomy and Space Science Major, Korea University of
Science and Technology, Daejeon 34113, Korea}

\affil{$^{17}$Max-Planck-Institute for Astronomy, K\"{o}nigstuhl 17,
69117 Heidelberg, Germany}

\affil{$^{18}$Department of Physics, Chungbuk National University,
Cheongju 28644, Republic of Korea}

\affil{$^{19}$School of Space Research, Kyung Hee University,
Yongin, Kyeonggi 17104, Korea}

\affil{$^{20}$IPAC, Mail Code 100-22, Caltech, 1200 E. California
Blvd., Pasadena, CA 91125, USA}

\affil{$^{21}$Dipartimento di Fisica ``E. R. Caianiello",
Universit\`{a} di Salerno, Via Giovanni Paolo II, 84084 Fisciano
(SA), Italy}

\affil{$^{22}$Spitzer, Science Center, MS 220-6, California
Institute of Technology,Pasadena, CA, USA}

\affil{$^{23}$NASA Exoplanet Science Institute, California Institute
of Technology, Pasadena, CA 91125, USA}

\affil{$^{24}$Institute for Space-Earth Environmental Research,
Nagoya University, Nagoya 464-8601, Japan}

\affil{$^{25}$Code 667, NASA Goddard Space Flight Center, Greenbelt,
MD 20771, USA; Email: david.bennett@nasa.gov}

\affil{$^{26}$Department of Physics, University of Auckland, Private
Bag 92019, Auckland, New Zealand}

\affil{$^{27}$Okayama Astrophysical Observatory, National
Astronomical Observatory of Japan, 3037-5 Honjo, Kamogata, Asakuchi,
Okayama 719-0232, Japan}

\affil{$^{28}$Department of Earth and Space Science, Graduate School
of Science, Osaka University, Toyonaka, Osaka 560-0043, Japan}

\affil{$^{29}$Nagano National College of Technology, Nagano
381-8550, Japan}

\affil{$^{30}$Tokyo Metropolitan College of Aeronautics, Tokyo
116-8523, Japan}

\affil{$^{31}$School of Chemical and Physical Sciences, Victoria
University, Wellington, New Zealand}

\affil{$^{32}$Institute of Space and Astronautical Science, Japan
Aerospace Exploration Agency, Kanagawa 252-5210, Japan}

\affil{$^{33}$University of Canterbury Mt.\ John Observatory, P.O.
Box 56, Lake Tekapo 8770, New Zealand}

\affil{$^{34}$Department of Physics, Faculty of Science, Kyoto
Sangyo University, 603-8555 Kyoto, Japan}

\affil{$^{35}$Kepler $\&$ K2 Missions, NASA Ames Research Center, PO
Box 1,M/S 244-30, Moffett Field, CA 94035}

\affil{$^{36}$Vanderbilt University, Department of Physics $\&$
Astronomy, Nashville, TN 37235, USA}

\affil{$^{37}$National Astronomical Observatories, Chinese Academy
of Sciences, A20 Datun Rd., Chaoyang District, Beijing 100012,
China}

\affil{$^{38}$Jodrell Bank Centre for Astrophysics, Alan Turing
Building, University of Manchester, Manchester M13 9PL, UK}

\affil{$^{39}$CFHT Corporation, 65-1238 Mamalahoa Hwy, Kamuela,
Hawaii 96743, USA}

\affil{$^{40}$Universit\'e de Toulouse, UPS-OMP, IRAP, Toulouse,
France}

\affil{$^{41}$Las Cumbres Observatory Global Telescope Network, 6740
Cortona Drive, suite 102, Goleta, CA 93117, USA}

\affil{$^{42}$Zentrum f{\"u}r Astronomie der Universit{\"a}t
Heidelberg, Astronomisches Rechen-Institut, M{\"o}nchhofstr. 12-14,
69120 Heidelberg, Germany}

\affil{$^{43}$Sorbonne Universit\'es, UPMC Univ Paris 6 et CNRS, UMR
7095, Institut d'Astrophysique de Paris, 98 bis bd Arago, 75014
Paris, France}

\affil{$^{44}$School of Physics and Astronomy, Tel Aviv University,
Tel Aviv 69978, Israel}

\affil{$^{45}$Planetary and Space Sciences, Department of Physical
Sciences, The Open University, Milton Keynes, MK7 6AA, UK}

\affil{$^{46}$Astrophysics Research Institute, Liverpool John Moores
University, Liverpool CH41 1LD, UK}

\affil{$^{47}$Istituto Nazionale di Fisica Nucleare, Sezione di
Napoli, Napoli, Italy}

\affil{$^{48}$Universit\"{a}t Hamburg, Faculty of Mathematics,
Informatics and Natural Sciences, Department of Earth Sciences,
Meteorological Institute, Bundesstra{\ss}e 55, 20146 Hamburg,
Germany}

\affil{$^{49}$Department of Astronomy, Stockholm University,
AlbaNova University Center, 106 91 Stockholm, Sweden}

\affil{$^{50}$INAF-Osservatorio Astronomico di Capodimonte, Salita
Moiariello 16, 80131, Napoli, Italy}

\affil{$^{51}$Astrophysics Group, Keele University, Staffordshire,
ST5 5BG, UK}

\affil{$^{52}$Max Planck Institute for Solar System Research,
Justus-von-Liebig-Weg 3, 37077 G{\"o}ttingen, Germany}

\affil{$^{53}$School of Physical Sciences, Faculty of Science,
Technology, Engineering and Mathematics, The Open University, Walton
Hall, Milton Keynes, MK7 6AA, UK}

\affil{$^{54}$Unidad de Astronom{\'{\i}}a, Fac. de Ciencias
B{\'a}sicas, Universidad de Antofagasta, Avda. U. de Antofagasta
02800, Antofagasta, Chile}

\affil{$^{55}$Instituto de Astrof\'isica, Pontificia Universidad
Cat\'olica de Chile, Av. Vicu\~na Mackenna 4860, 7820436 Macul,
Santiago, Chile}

\affil{$^{56}$Department of Physics, Sharif University of
Technology, PO Box 11155-9161 Tehran, Iran}

\affil{$^{57}$Department of Physics, Isfahan University of
Technology, 841568311 Isfahan, Iran}

\affil{$^{58}$Centre for Electronic Imaging, Department of Physical
Sciences, The Open University, Milton Keynes, MK7 6AA, UK}

\affil{$^{59}$Stellar Astrophysics Centre, Department of Physics and
Astronomy, Aarhus University, Ny Munkegade 120, 8000 Aarhus C,
Denmark}

\affil{$^{\dag}$NASA Postdoctoral Program Fellow}

\affil{$^{\dag\dag}$Sagan Fellow}

\affil{$^{\dag\dag\dag}$Royal Society University Research Fellow}





\begin{abstract}

We report the discovery of OGLE-2016-BLG-1190Lb, which is likely to be
the first {\it Spitzer} microlensing planet in the Galactic bulge/bar,
an assignation that can be confirmed by two epochs of high-resolution
imaging of the combined source-lens baseline object.  The planet's
mass $M_p= 13.4\pm 0.9\,M_J$ places it right at the deuterium burning
limit, i.e., the conventional boundary between ``planets'' and ``brown
dwarfs''.  Its existence raises the question of whether such objects
are really ``planets'' (formed within the disks of their hosts) or
``failed stars'' (low mass objects formed by gas fragmentation).
This question may ultimately be addressed by comparing disk and
bulge/bar planets, which is a goal of the {\it Spitzer} microlens
program.  The host is a G dwarf $M_{\rm host} = 0.89\pm
0.07\,M_\odot$ and the planet has a semi-major axis $a\sim 2.0\,\au$.
We use {\it Kepler} K2 Campaign 9 microlensing data to break
the lens-mass degeneracy that generically impacts parallax solutions
from Earth-{\it Spitzer} observations alone, which is the first
successful application of this approach.
The microlensing data, derived primarily from near-continuous,
ultra-dense survey observations from OGLE, MOA, and three KMTNet
telescopes, contain more orbital information than
for any previous microlensing planet, but not quite enough to
accurately specify the full orbit.  However, these data do
permit the first rigorous test of microlensing orbital-motion
measurements, which are typically derived from data taken over $<1\%$
of an orbital period.  

\end{abstract}


\section{{Introduction}
\label{sec:intro}}

The discovery of {\it Spitzer} microlensing planet
OGLE-2016-BLG-1190Lb is remarkable in five different respects.  First,
it is the first planet in the {\it Spitzer} Galactic-distribution
sample that likely lies in the Galactic bulge, which would break the
trend from the three previous members of this sample.  Second, it is
precisely measured to be
right at the edge of the brown dwarf desert.  Since the existence of
the brown dwarf desert is the signature of different formation
mechanisms for stars and planets, the extremely close proximity of
OGLE-2016-BLG-1190Lb to this desert raises the question of whether it
is truly a ``planet'' (by formation mechanism) and therefore reacts
back upon its role tracing the Galactic distribution of planets, just mentioned
above.  Third, it is the first planet to enter the {\it Spitzer}
``blind'' sample whose existence was recognized prior to its choice as
a {\it Spitzer} target.  This seeming contradiction was clearly
anticipated by \citet{yee15} when they established their protocols for
the Galactic distribution experiment.  The discovery therefore tests
the well-defined, but intricate procedures devised by \citet{yee15} to
deal with this possibility.  Fourth, it is the first planet (and indeed
the first microlensing event) for which the well-known microlens-parallax
degeneracy has been broken by observations from two satellites.
Finally, it is the first microlensing
planet for which a complete orbital solution has been attempted.
While this attempt is not completely successful in that a
one-dimensional degeneracy remains, it is an important benchmark on
the road to such solutions.

In view of the diverse origins and implications of this discovery, we
therefore depart from the traditional form of introductions and begin
by framing this discovery with four semi-autonomous introductory
subsections.

{\subsection{Microlens Parallax from One and Two Satellites}
\label{sec:k2_intro}}

When \citet{refsdal66} first proposed to measure microlens parallaxes
using a satellite in solar orbit, a quarter century before the
first microlensing event, he already realized that this measurement
would be subject to a four-fold degeneracy, and further, that this
degeneracy could be broken by observations from a second satellite.
See also \citet{gould94b} and \citet{scn16}.
The microlens parallax is a vector
\begin{equation}
\bpi_\e \equiv \pi_\e{\bmu_\rel\over\mu_\rel};
\qquad \pi_\e \equiv {\pi_\rel\over\theta_\e},
\label{eqn:ulenspar}
\end{equation}
whose amplitude is the ratio of the lens-source relative parallax
$\pi_\rel=\au(D_L^{-1} - D_S^{-1})$ to the Einstein radius
$\theta_\e$, and whose direction is that of the lens-source relative
proper motion $\bmu_\rel$.  As illustrated by Figure~1 of
\citet{gould94b} (compare to Figure~1 of \citealt{yee15a}) observers
from Earth and a satellite will see substantially different light
curves.  By comparing the two light curves, one can infer the
vector offset within the Einstein ring of the source as seen from
the two observers.  Combining this vector offset with the known
projected offset of the satellite and Earth, one can then infer $\bpi_\e$.

However, this determination is in general subject to a four-fold
degeneracy.  While the component of the vector offset in the direction
of lens-source motion $\bmu_\rel$ gives rise to an offset in peak times
of the event and can therefore be determined unambiguously, the
component transverse to this motion must be derived from a comparison
of the impact parameters, which leads to a four-fold ambiguity.
That is, the impact parameter is a signed quantity but only its magnitude
can be readily determined from the light curve.

By far, the most important aspect of this degeneracy is that
the source may be either on the same or opposite sides of
lens as seen from the two observatories.  The parallax amplitude
$\pi_\e$ will be smaller in the first case than the second,
which will directly affect the derived lens mass $M$ and
$\pi_\rel$ \citep{gould92,gould04}
\begin{equation}
M = {\theta_\e\over \kappa\pi_\rel};
\qquad
\pi_\rel =\pi_\e\theta_\e;
\qquad
\kappa \equiv {4 G\over c^2\au}\simeq 8.1\,{\mas\over M_\odot} .
\label{eqn:mpirel}
\end{equation}
By contrast, the remaining two-fold degeneracy only impacts the
inferred direction of motion, which is usually of little physical
interest.

The first such parallax measurement was made by \citet{smc001} by
combining {\it Spitzer} and ground based observations of
OGLE-2005-SMC-001, toward the Small Magellanic Cloud.
Subsequently, more than 200 events were observed toward the Galactic
bulge in 2014 and 2015 as part of a multi-year {\it Spitzer} program
\citep{prop2013,prop2014} of which more than 70 have already been
published.  A key issue in the analysis of these events has been to
break this four-fold degeneracy, in particular the two-fold
degeneracy that impacts the mass and distance estimates. While in
some cases \citep{yee15a,ob141050}, this degeneracy has been broken
by various fairly weak effects, in the great majority of cases, the
degeneracy was broken only statistically \citep{21event,zhu17}.

While such statistical arguments are completely adequate when the
derived conclusions are themselves statistical, they are less
satisfactory for drawing conclusions about individual objects.
Hence, for the 2016 season, \citet{prop2015b} specifically proposed
to observe some events with {\it Spitzer} that lay in the roughly
$4\,{\rm deg}^2$ observed by {\it Kepler} during its K2 Campaign 9,
in addition to the regular {\it Spitzer} targets drawn from a much
larger $\sim 100\,{\rm deg}^2$ area \citep{prop2015a}.  Contrary to
the expectations of \citet{refsdal66} and \citep{gould94b}, {\it
Spitzer}, {\it Kepler}, Earth, and the microlensing fields all lie
very close to the ecliptic, so that the projected positions of the
sources as seen from the three observatories are almost colinear.
This means that it is almost impossible to use {\it Kepler} to fully
break the four-fold degeneracy.  Nevertheless, this configuration
does not adversely impact {\it Kepler}'s ability to break the key
two-fold degeneracy that impacts the mass and distance
determinations, which turns out to be quite important in the present
case. (See also \citealt{zhu17c} for the case of a single-lens
event.)

{\subsection{Planets at the Desert's Edge}
\label{sec:desert_intro}}

The term ``brown dwarf desert'' was originally coined by
\citet{marcy00} to describe the low frequency of ``brown dwarfs'' in
Doppler (RV) studies relative to ``planets'' of somewhat lower mass.
Since the sensitivity of the surveys rises with mass, this difference
cannot be due to selection effects.
Later, \citet{grether06} quantified this desert as the intersection of
two divergent power laws, subsequently measured as $d N/d\ln M \sim
M^{-0.3}$ for ``planets'' and $d N/d\ln M \sim M^{1}$ for ``stars''.
We have placed all these terms in quotation marks because they are
subject to three different definition systems that are not wholly
self-consistent.  By one definition system, planets, brown dwarfs, and
stars are divided by mass at $13\,M_{\rm J}$ and $0.08\,M_\odot$.  By a second
they are divided at deuterium and hydrogen burning.  And in a third
system they are divided by formation mechanism: in-disk formation for
planets, gravitational collapse for stars, and [either or both] for
brown dwarfs.

The first definition has the advantage that mass is something that can
in principle be measured.  The second system is valuable because it
permits a veneer of physical motivation on what is actually an
arbitrary boundary.  In fact, no plausible mechanism has ever been
advanced as to how either deuterium burning or hydrogen burning can
have any impact on the mass function of the objects being formed.  In
particular, hydrogen burning commences in very low mass stars long
after they have become isolated from their sources of mass accretion.
The third definition speaks to a central scientific question about
these various types of objects: where do they come from?
Unfortunately, for field objects, there is precious little
observational evidence that bears on this question.  Up until now, the
key input from observations is statistical: far from the boundaries,
planets and stars follow divergent power laws, which almost certainly
reflect different formation mechanisms \citep{grether06}.  However,
near the boundary, in particular in the brown dwarf desert and on its
margins, there is no present way to map individual objects onto a
formation mechanism even if their masses were known.  Moreover, using
the RV technique, i.e., the traditional method for finding brown dwarf
companions at few AU separations, there is no way to precisely measure
the masses (unless, by extreme chance, the system happens to be eclipsing).

If the divergent power laws (as measured well away from their
boundaries) represent different formation mechanisms, then most likely
these power laws continue up to and past these nominal boundaries, so
that ``brown dwarfs'' as defined by mass
represent a mixture of populations as defined by
formation, and high-mass ``planets'' do as well.

Microlensing opens several different laboratories for disentangling
formation mechanism from mass, at least statistically.  First, as
pointed out by \citet{mb07197} and \citet{ob160693}, microlensing
can probe to larger orbital radii than RV for both massive planets
and brown dwarfs and so determine whether the independent mode of
planet formation ``dies off'' at these radii and, if so, how this
correlates to the behavior of brown dwarfs.  Second, it can probe
seamlessly to the lowest-mass hosts of brown dwarfs, even into the
brown dwarf regime itself.  This is a regime that is progressively
less capable of forming brown dwarfs from disk material, although it
may be proficient at forming Earth-mass planets \citep{ob161195}.
Third, since microlensing is most directly sensitive to the
companion/host mass ratio $q$, it can precisely measure the
distribution of this parameter, even for samples for which the
individual masses are poorly known\footnote{As a result, in
microlensing statistical studies, the planet/BD boundary is often
defined by $q$.  For example, \citet{suzuki16} (following
\citealt{ob03235}) and \citet{shvartz16} use $q=0.03$ and $q=0.04$,
respectively, which would correspond to the conventional $13\,M_{\rm
jup}$ limit for stars of mass $M\simeq 0.4\,M_\odot$ and $M\simeq
0.3\,M_\odot$, respectively.}. The minimum in this distribution can
then be regarded as the location of the mean boundary between two
formation mechanisms averaged over the microlensing host-mass
distribution. \citet{shvartz16} found that this minimum was near
$q\sim 0.01$, which corresponds to $M_{\rm comp}\sim 5\,M_{\rm Jup}$
for characteristic microlensing hosts, which are typically in the M
dwarf regime.  This tends to indicate that this boundary scales as a
function of the host mass.

Another path open to microlensing is probing radically different
star-forming environments, in particular the Galactic bulge.
\citet{thompson13}, for example, has suggested that massive-planet
formation via the core-accretion scenario was strongly suppressed in
the Galactic bulge by the high-radiation environment.  This would not
impact rocky planets but would lead to a dearth of Jovian planets and
super-Jupiters if these indeed formed by this mechanism.  Of
particular note in this regard is that adaptive optics observations by
\citet{mb11293b} indicated that MOA-2011-BLG-293Lb \citep{mb11293} is
a $5\,M_J$ object orbiting a solar-like host in the Galactic bulge.
This might be taken as evidence against Thompson's conjecture.
However, another possibility is that MOA-2011-BLG-293Lb formed at the
low-mass end of the gravitational-collapse mode that produces most
stars, which was perhaps
more efficient in the high-density, high-radiation environment that
characterized early star formation in the bulge.  In this case, we
would expect the companion mass function in the Galactic bulge
to be rising toward the deuterium-burning limit, in sharp contrast to the
mass function in the Solar-neighborhood, which is falling in this range.
That is, high-mass planets (near the deuterium-burning limit) would
be even more common than the super-Jupiter found by \citet{mb11293b}.

{\subsection{Construction of Blind Tests In the Face of ``Too Much'' Knowledge}
\label{sec:yee_intro}}

\citet{yee15} proposed to measure the Galactic distribution of planets
by determining individual distances to planetary (and non-planetary)
microlenses from the combined analysis of light curves obtained from
ground-based and {\it Spitzer} telescopes.  Because the lenses are
usually not directly detected, such distance measurements are
relatively rare in the absence of space-based microlens parallax
\citep{refsdal66,gould94b} and, what is more important, heavily biased
toward nearby lenses.

As \citet{yee15} discuss in considerable detail, it is by no means
trivial to assemble a {\it Spitzer} microlens-parallax sample
\citep{prop2013,prop2014,prop2015a,prop2015b,prop2016} that is
unbiased with respect to the presence or absence of planets.
\citet{21event} showed how the cumulative distribution of planetary
events as a function of distance toward the Galactic bulge could be
compared to that of the parent sample to determine whether planets are
relatively more frequent in the Galactic disk or bulge.  However, this
comparison depends on the implicit assumption that there is no bias
toward selection of planetary events.  In fact, it would not matter if
the planetary sample were biased, provided that this bias were equal
for planets in both the Galactic disk and bulge.  However,
particularly given the constraints on {\it Spitzer}
target-of-opportunity (ToO) selection, it is essentially impossible to
ensure such a uniform bias without removing this bias altogether.

Hence, \citet{yee15} developed highly articulated protocols for
selecting {\it Spitzer} microlens targets that would ensure that the
resulting sample was unbiased.  We will review these procedures in
some detail in Section~\ref{sec:yee_proc}.  However, from the present
standpoint the key point is that however exactly the sample is
constructed, it must contain only events with ``adequately measured''
microlens parallaxes.  \citet{yee15} did not specify what was
``adequate'' because this requires the study of real data.
\citet{zhu17} carried out such a study based on a sample of 41 {\it
  Spitzer} microlensing events without planets, which meant that these
authors could not be biased -- even unconsciously -- by a ``desire''
to get more planets into the sample.  In addition, they specifically
did not investigate how their criteria applied to the two {\it
  Spitzer} microlens planets that were previously discovered
\citep{ob140124,ob150966} until after these criteria were decided.
The \citet{zhu17} criteria, as they apply to non-planetary events, are
quite easy to state once the appropriate definitions are in place
(Section~\ref{sec:zhu_proc}).  A crucial point, however, is that for
planetary events, these same criteria must be applied to the
point-lens event that would have been observed in the absence of
planets.

Thus, while in some cases, it may be quite obvious whether a planetary
event should or should not be included in the sample, it is also
possible that this assignment may require rather detailed analysis.

The {\it Spitzer} microlens planetary event OGLE-2016-BLG-1190 does in
fact require quite detailed analysis to determine whether it belongs
in the {\it Spitzer} Galactic distribution of planets sample.
OGLE-2016-BLG-1190 was initially chosen for {\it Spitzer} observations
based solely on the fact that it had an anomaly that was strongly
suspected to be (and was finally confirmed as) planetary in nature.
At first sight, this would seem to preclude its participation in an
``unbiased sample''.  Nevertheless, \citet{yee15} had anticipated this
situation and developed protocols that enable, under some
circumstances, the inclusion of such planets without biasing the
sample.  We show that OGLE-2016-BLG-1190 in fact should be included
under these protocols.  This then sets the stage for whether its
parallax is ``adequately measured'' according to the \citet{zhu17}
criteria, or rather whether the corresponding point-lens event would
have satisfied them.  We address this point for the first time here as
well.

{\subsection{Full Kepler Orbits in Microlensing}
\label{sec:kepler_intro}}

When microlensing planet searches were first proposed
\citep{mao91,gouldloeb}, it was anticipated that only the planet-star
mass ratio $q$ and projected separation (scaled to the Einstein radius
$\theta_\e$) $s$, would be measured.  Even the mass $M$ of the host
was thought to be subject only to statistical estimates, while orbital
motion was not even considered.  It was quickly realized, however,
that it was at least in principle possible to measure both $\theta_\e$
\citep{gould94a} and the microlens parallax $\pi_\e$ \citep{gould92}
\begin{equation}
\theta_\e^2\equiv \kappa M\pi_\rel;
\qquad
\pi_\e^2\equiv {\pi_\rel\over\kappa M};
\qquad
\kappa\equiv {4G\over c^2\au}\simeq 8.14{\mas\over M_\odot},
\label{eqn:thetaepie}
\end{equation}
and that this could then yield both the lens mass
$M=\theta_\e/\kappa\pi_\e$ and the lens-source relative parallax
$\pi_\rel=\theta_\e\pi_\e$.

The fact that linearized orbital motion was measurable was discovered
by accident during the analysis of the binary microlensing event
MACHO-97-BLG-41 \citep{mb9741}.  In a case remarkably similar to the
one we will be analyzing here, the source first passed over an
outlying caustic of a close binary and later went over the central
caustic.  From the analysis of the latter, one could determine $(s,q)$
and ``predict'' the positions of the two outlying caustics.  These
differed in both coordinates from the caustic transit that had
actually been observed.  The difference was explained in terms of
binary orbital motion, and the linearized orbital parameters were thus
measured.  This was regarded at the time as requiring very special
geometry because the typical duration of caustic-induced effects is a
few days whereas the orbital period of systems probed by microlensing
is typically several years.  In fact, however, orbital motion began to
be measured or constrained in many planetary events, mostly with quite
generic geometries, including the second microlensing planet
OGLE-2005-BLG-071Lb \citep{ob05071,ob05071b}.  A fundamental feature
of microlensing that enables such measurements is that the times of
caustic transits can often be measured with precisions of better than
one minute.  Still, it did not seem possible to measure full orbits.
Nevertheless, \citet{ob09020} significantly constrained all 7 Kepler
parameters for the binary system OGLE-2009-BLG-020L, albeit with huge
errors and strong correlations. These measurements
were later shown to be consistent with RV followup observations
by \citet{yee16}.  Subsequently, \citet{ob05018,ob110417}
fully measured all Kepler parameters for several different binaries.

To date, and with one notable exception, such complete Kepler
solutions have been more of interest in terms of establishing the
principles and methods of making the measurements than anything they
are telling us about nature.  The exception is OGLE-2006-BLG-109La,b,
the first two planet system found by microlensing
\citep{ob06109,ob06109b}.  Due to the very large caustic from one of
the planets, together with a data rate that was very high and
continuous for that time,  \citet{ob06109b} were able to introduce
one additional dynamical parameter relative to the standard
two-dynamical parameter approach of \citet{mb9741}.  This
allowed them make RV predictions for the system that could be
tested with future 30m class telescopes.

However, if the method of measuring complete Kepler orbits can be extended
from binaries to planets (as we begin to
do here) then it will permit much stricter comparison between RV and
microlensing samples, which has so far been possible only
statistically, (e.g., \citealt{gould10,clanton14,clanton14b,clanton16}).
In particular, we provide here the first evidence for a non-circular
orbit of a microlensing planet.

\section{{Observations}
\label{sec:obs}}

\subsection{{Ground-Based Observations}
\label{sec:groundobs}}

OGLE-2016-BLG-1190 is at (RA,Dec) = (17:58:52.30,$-27$:36:48.8),
corresponding to $(l,b)=(2.63,-1.84)$.
It was discovered and announced as a probable microlensing event
by the Optical Gravitational Lensing Experiment (OGLE) Early Warning
System \citep{ews1,ews2} at UT 17:37 on 27 June 2016.  OGLE observations
were at a cadence of $\Gamma = 3\,{\rm hr^{-1}}$ using their 1.3m telescope
at Las Campanas, Chile.

The event was independently
discovered on 6 July by the Microlensing Observations in Astrophysics (MOA)
collaboration
as MOA-2016-BLG-383 based on $\Gamma = 4\,{\rm hr^{-1}}$ observations
using their 1.8m telescope at Mt. John, New Zealand.

The Korea Microlensing
Telescope Network (KMTNet, \citealt{kmtnet})
observed this field from its three 1.6m
telescopes at CTIO (Chile), SAAO (South Africa) and SSO (Australia),
in its two slightly offset fields BLG03 and BLG43, with combined
cadence of $\Gamma = 4\,{\rm hr^{-1}}$.

The great majority of observations were in $I$ band for OGLE and KMTNet and
a broad $RI$ band for MOA, with occasional $V$ band observations made
solely to determine source colors.  All reductions for the light curve
analysis were conducted using variants of difference image analysis (DIA,
\citealt{alard98}), specifically \citet{wozniak2000} and \citet{albrow09}.

In addition to these high-cadence, near-continuous survey
observations, OGLE-2016-BLG-1190 was observed in two lower-cadence
surveys that were specifically motivated to support microlensing in
the {\it Kepler} microlensing (K2C9) field \citep{henderson16}, in
which it lies.  These surveys, respectively by the 3.6m
Canada-France-Hawaii Telescope (CFHT) and the 3.8m United Kingdom
Infrared Telescope (UKIRT) are both located at the Mauna Kea
Observatory in Hawaii.
The CFHT observations were carried out equally in $g$, $r$, and $i$
bands, but only the latter two are incorporated in the fit because the
$g$ data are too noisy.  The UKIRT observations were in $H$-band;
these are used here solely for the purpose of measuring the $H$-band
source flux, and so the $(I-H)_s$ source color.

Finally, two follow-up groups started to monitor the event shortly
after the public announcement (just before peak) of an anomaly by the
MOA group.  These were RoboNet and MiNDSTEp.  Both observatories began
observing immediately, i.e., just after peak, from SAAO using the
LCO 1m and from the Danish 1.5m at La Silla, Chile, respectively.

In the latter case the observations were triggered automatically by
the SIGNALMEN algorithm \citep{signalman} after it detected an
anomaly at  ${\rm HJD}^\prime=7581.0$, with the observations
themselves beginning 0.73 days later.  The observations were
taken by the EMCCD camera at 10 Hz \citep{emccd} in $V$ and $I$, but
only the $I$ band data were used in the analysis.

\subsection{{{\it Spitzer} Observations}
\label{sec:spitzerobs}}

At UT 02:44, 29 June, YHR sent a message to the {\it Spitzer} team
reporting his ``by eye'' detection of an anomaly at HJD$^\prime(\equiv
{\rm HJD}-2450000)\sim 7500$, which he had tentatively modeled as
being due to a brown-dwarf (BD) or planetary companion.  That is, the
putative anomaly had occurred about 69 days previously, and indeed 67
days before the OGLE alert.  Since this anomaly alert was also 12 days
before peak, when the event was only 0.3 mag above baseline, it was
impossible at that time to accurately estimate the basic parameters of
the event.  Based on this alert (and subsequent additional modeling
using KMTNet data), the {\it Spitzer} team initiated regular cadence
($\Gamma \sim 1\,{\rm day^{-1}}$) observations at the next
upload, leading to a total of 19 observations during $7578<{\rm
  HJD}^\prime<7596$.  The data were reduced using specially designed
software for microlensing \citep{170event}.
We note that it was the promptness of the OGLE
alert that enabled recognition of the much earlier anomaly in time to
trigger {\it Spitzer} observations over the peak of the event.

Table~\ref{tab:observe} specifies the number of data points and
filter(s) of each observatory, as well as its contribution to the
total $\chi^2$ of the best model (described in Section~\ref{sec:kepler}).

\section{{Analysis}
\label{sec:analysis}}

\begin{figure}
\plotone{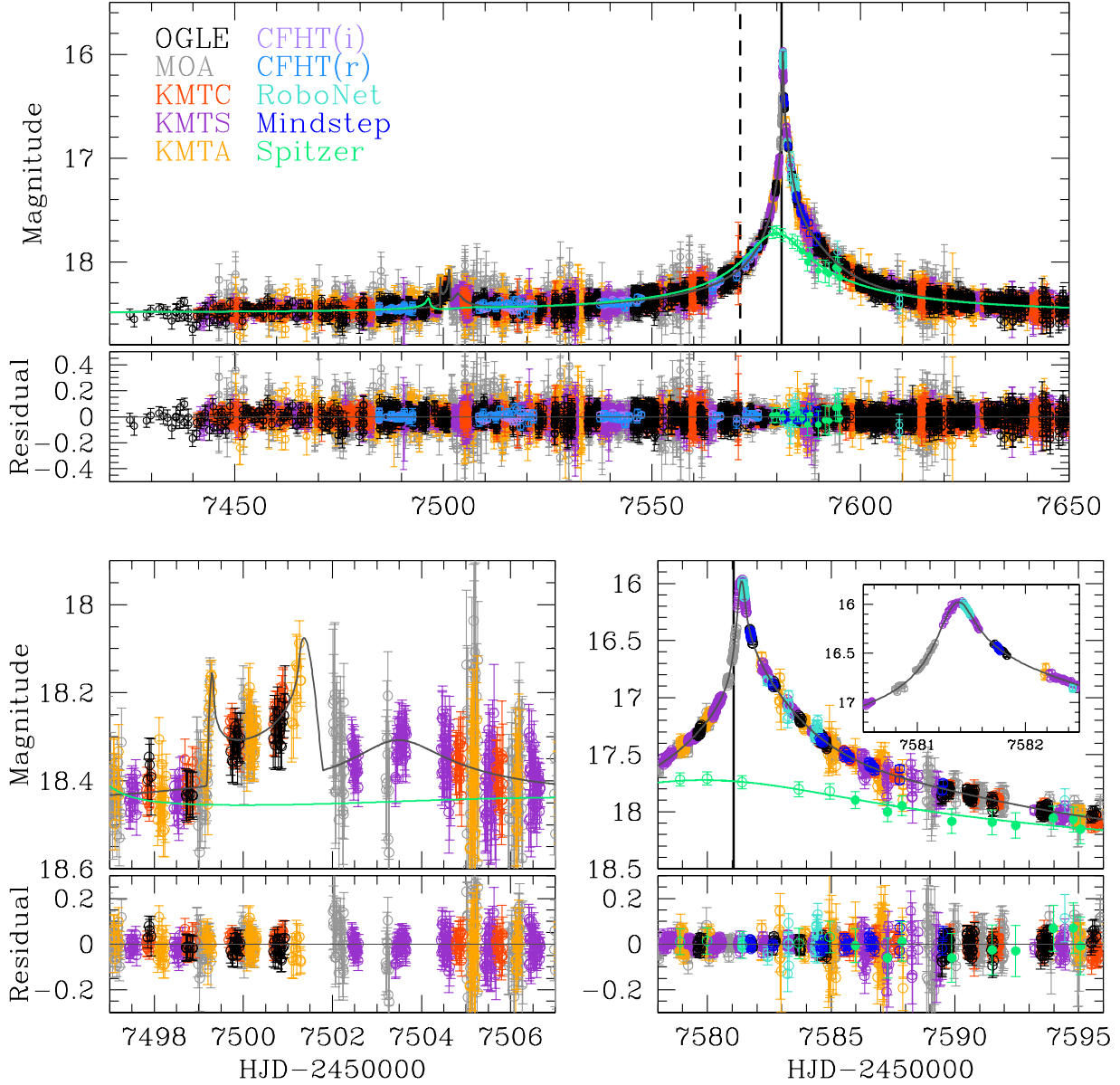}
\caption{Light curve of OGLE-2016-BLG-1190.  The data points
are colored as indicated by observatory
in the top panel, which shows the full light curve.
Fluxes $f_i$ from observatory $i$ (including {\it Spitzer})
are aligned to the OGLE scale by
$f^\prime_{i,\rm obs} = (f_{i,\rm obs} - f_{b,i})(f_{s,\rm ogle}/f_{s,i})+f_{b,\rm ogle}$.
Models are shown for ground-based and {\it Spitzer} data in black and
green, respectively.  Vertical dashed and solid lines indicate
the subjective and objective selection dates for {\it Spitzer} observations,
respectively. Open and filled circles for {\it Spitzer} data (green)
show observations initiated by the subjective and objective selection,
respectively.
Lower panels show zooms of the planetary-caustic crossing (left) and
central-caustic cusp approach (right).
}
\label{fig:lc}
\end{figure}

\begin{figure}
\plotone{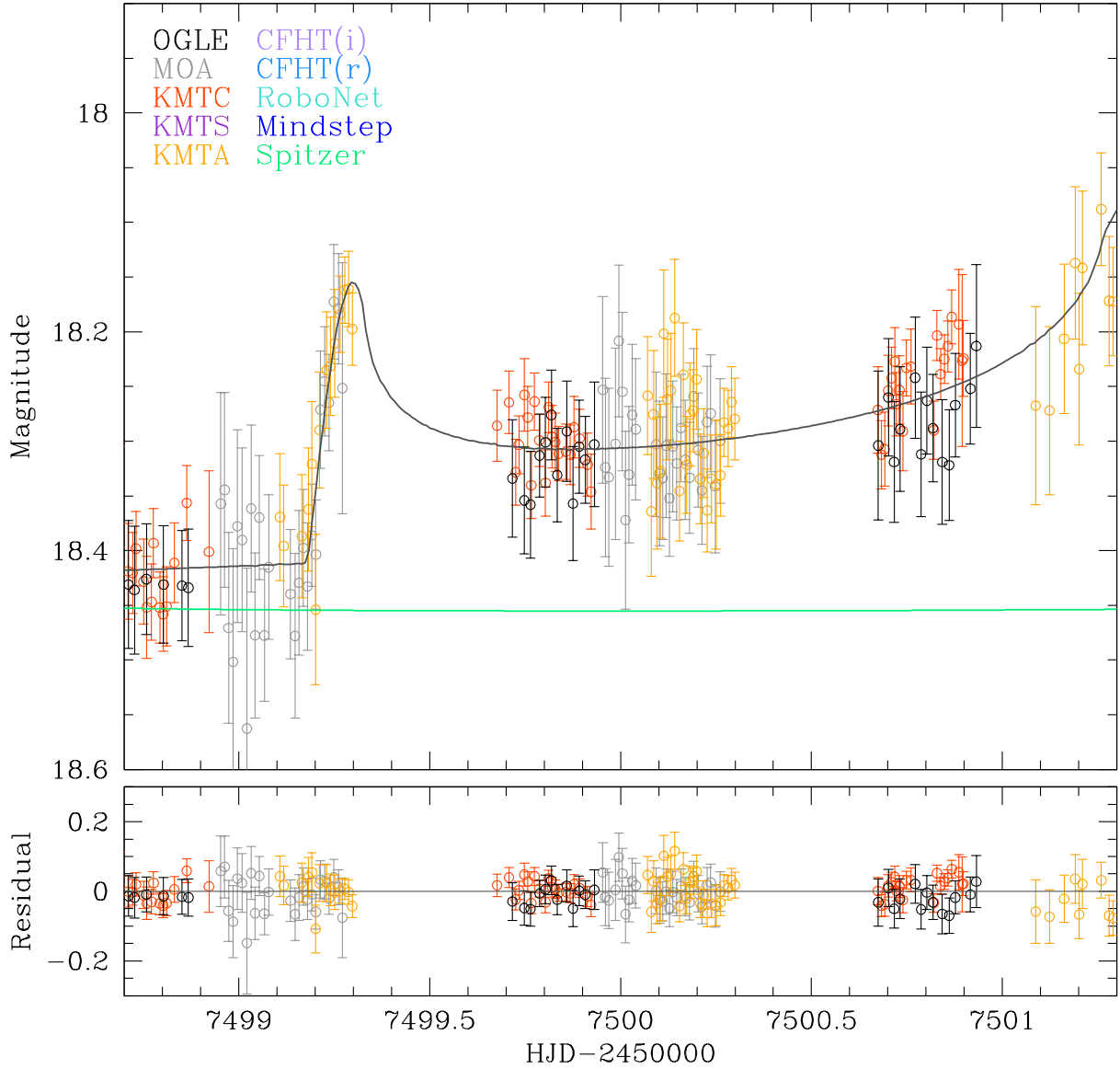}
\caption{Further zoom of the lower-left panel of Figure~\ref{fig:lc},
focusing on the data approaching and within the planetary caustic.
The caustic entrance is well-defined by the KMTA and MOA
data, with residuals that are consistent with the errors and that
show no significant systematic trends.
}
\label{fig:lc_zoom}
\end{figure}

Figures~\ref{fig:lc} and \ref{fig:lc_zoom}
show the light curve of all the data together with
a best fitting model.  Ignoring the model for the moment, the data show
two clear isolated deviations from a smooth point-lens model: an
irregular bump at HJD$^{\prime}\sim 7500$ and an asymmetric peak at
HJD$^{\prime} \sim 7581$.  Figure~\ref{fig:lc} shows the
overall light curve together with two zooms featuring the regions
around the these two anomalies, while  Figure \ref{fig:lc_zoom}
shows a further zoom of the first anomaly.
In addition, the data from the {\em Spitzer}
spacecraft show a clear parallax effect, i.e., although the data are
taken contemporaneously with the ground-based data, the light curves
observed from the two locations are clearly different. The final model
for the light curve must account for all of these effects: the two
deviations from the point lens and the parallax effect seen from {\em
  Spitzer}.

The nature of the two deviations can be understood through the
ground-based data alone. The two deviations could be caused by the
same planet or, in principle, by two different companions to the host
star. As we will show in Section \ref{sec:groundmod}, a single planet
that explains the central caustic perturbation at HJD$^{\prime} \sim
7581$ actually predicts the existence of the planetary caustic
perturbation at HJD$^{\prime}\sim 7500$ if the source trajectory is
slightly curved. Such a curvature implies that we observe the orbital
motion of the planet, and since orbital motion is partially degenerate
with the parallax effect, in Section~\ref{sec:linearorb} we proceed
with fitting the ground-based and {\em Spitzer} data together with
both effects. In that section, we show that the prediction of the
planetary caustic crossing is remarkably precise. Thus, for our final
fits in Section~\ref{sec:kepler}, we model the light curve using a
full Keplerian prescription for the orbit.

\subsection{{Ground-Based Model}
\label{sec:groundmod}}

The simplest explanation for the ground-based lightcurve is that both
deviations could be due to a single companion. All companions that are
sufficiently far from the Einstein ring produce two such sets of
caustics, one set of outlying ``planetary'' caustic(s) and one
``central'' caustic.  For wide-separation companions $(s>1)$, the
second caustic lies directly on the binary axis.  For close companions
$(s<1)$ there are two caustics that are symmetric about this axis, but
for low-mass companions, $(q\ll 1)$, these caustics lie close to the
binary axis.  Thus, a planetary companion can generate two
well-separated deviations provided that the angle of the source
trajectory relative to the binary axis satisfies $\alpha\sim 0$ or
$\alpha\sim\pi$.  If this is the true explanation, then the central
caustic crossing should be consistent with a source traveling
approximately along the binary axis of that caustic. If the central
caustic crossing is not consistent with such a configuration, e.g., it
would require a source traveling perpendicular to the binary axis,
that would be evidence that the two deviations were due to two
separate companions. To test whether there is any evidence for the
latter hypothesis, we begin by excising the data from the isolated,
first anomaly and fitting the rest of the ground-based light curve.

Such binary lens fits require a minimum of six parameters
$(t_0,u_0,t_\e,s,q,\alpha)$.  The first three are the standard
point-lens parameters \citep{pac86}, i.e., the time of lens-source
closest approach, the impact parameter normalized to the angular
Einstein radius $\theta_\e$, and the Einstein crossing time,
\begin{equation}
t_\e\equiv {\theta_\e\over\mu_\rel},
\label{eqn:tedef}
\end{equation}
where $\bmu_\rel$ is the lens-source relative proper motion
and $\mu_\rel=|\bmu_\rel|$.  While for
point lenses the natural reference point for $(t_0,u_0)$ is the
(single) lens, for binary lenses it must be specified.  We choose the
so-called center of magnification \citep{ob04343,mb07400}.  The
remaining three parameters are the companion-star separation $s$
(normalized to $\theta_\e$), their mass ratio $q$, and the angle
$\alpha$ of their orientation on the sky relative to $\bmu_\rel$.  If
the source comes close to or passes over the caustics, then one also needs
to specify $\rho\equiv\theta_*/\theta_\e$, where $\theta_*$ is the
source angular radius.  We note that for $s<1$, the center of
magnification is conveniently the same as the center of mass.

We model the light curve using inverse ray shooting
\citep{kayser86,schneider88,wambs97} when the source is close to a
caustic and multipole approximations \citep{pejcha09,gould08}
otherwise.  We initially consider an $(s,q,\alpha)$ grid of starting
points for Markov Chain Monte Carlo (MCMC) searches, with the
remaining parameters starting at values consistent with a point-lens
model.  Then $(s,q)$ are held fixed while the remaining parameters are
allowed to vary in the chain.  We then start new chains at each of the
local minima in the $(s,q)$ $\chi^2$ surface, with all parameters
allowed to vary.

We find the light curve excluding the early caustic crossing data can
be explained by a planet with parameters:
\begin{equation}
(s,q,\alpha)=(0.60,0.016,-0.01).
\label{eqn:sqalpha}
\end{equation}
As expected for a light curve generated by a single, low-mass
companion, $\alpha$ is indeed close to zero. For such
central-caustic events, we usually expect two solutions related by
the well-known close-wide $(s\leftrightarrow s^{-1})$ degeneracy
\citep{griest98,dominik99}. Thus, we might also expect a second
solution with parameters $(s,q,\alpha) = (1.67,0.016,-0.01)$.
However, although the central caustics of both the $s<1$ and $s>1$
solutions are quite similar, the planetary caustic lies on the
opposite side of the host as the planet for $s<1$ and on the same
side for $s>1$. As a result, because $\alpha\sim0$, the $s>1$
solution would produce a large planetary caustic crossing {\it
after} the central caustic crossing, which we do not observe.
Therefore, the $s<1$ solution is the only one that can explain the
light curve.

Nevertheless, at first sight it does not appear that the $s<1$
solution can explain the planetary caustic crossing at HJD$^{\prime}
\sim 7500$. The $s<1$ caustic geometry is characterized by two
caustics on opposite sides of this axis.  For $(s,q)=(0.6,0.016)$, the
angle between each caustic and the binary axis is $\phi\sim 16^\circ$
(see Equation~(\ref{eqn:xietacaust})).  Thus, given that the source
trajectory is very close to the planet-star axis ($|\alpha| \ll 1$),
it would appear that the source would pass between the two caustics
(e.g., the source travels along the $x$-axis between the red caustics
in Figure \ref{fig:geolin}), whereas we clearly see in the data
(Figure~\ref{fig:lc}) that the source must pass over a caustic at
$\sim 7500$.

However, this apparent contradiction can be resolved if the planet
(and so caustics) have moved during the $\sim 80$ days between the
times of the first perturbation and the second (when this geometry is
determined). Naively, we expect motion of order $d\alpha/dt \sim
17^\circ/(80\, {\rm days}) \sim 0.2^{\circ}{\rm day}^{-1}$.  This kind
of motion was indeed the resolution of the first such puzzle for
MACHO-97-BLG-41 \citep{mb9741x,mb9741,jung13}.

Hence, we conclude that the two perturbations are likely caused by a
single companion, with the proviso that we must still check that the
form of the planetary-caustic perturbation ``predicted'' by the
central caustic crossing is consistent with the observed perturbation
and that the amplitude of internal motion is consistent with a
gravitationally bound system.

\subsection{{Linearized Orbital Motion and the Microlens Parallax}
\label{sec:linearorb}}

Given our basic understanding of the anomaly from the ground-based
data, we can proceed with modeling the full dataset including {\it
  Spitzer} data. The ground-based modeling implies that the orbital
motion effect plays a prominent role, so we allow for linearized
motion of the lens system, i.e., we add two variables corresponding to
the velocity of the lens projected onto the plane of the sky,
$d\alpha/dt$ and $ds/dt$. Including the {\it Spitzer} data requires
also including the parallax effect. The combination of these two
effects implies the possibility of up to eight degenerate solutions:
two solutions because with orbital motion the source can pass through
either planetary caustic, multiplied by four solutions due to the
well-known satellite parallax degeneracy \citep{refsdal66,gould94b}.

We begin by describing the color-magnitude diagram (CMD) analysis in
Section \ref{sec:cmd} because it is used to derive the color-color
relation needed for combining the {\it Spitzer} and ground-based
data. Then, we give a qualitative evaluation of the {\it Spitzer}
parallax in Section \ref{sec:spitzpar}. In this section, we show that
the color-color constraint plays an important role in measuring the
parallax even though the {\it Spitzer} light curve partially captures
the peak of the event. In Section \ref{sec:modelorb}, we present the
full model of the event including linearized orbital motion and
parallax. This modeling demonstrates that the
orbital-motion parameters that are derived after excluding the $\pm 10\,$ days
of data around the planetary caustic crossing are very similar to those
derived from the full data set.  Furthermore, the information from this
restricted data set eliminates one of the two possible directions
of orbital motion.  Finally, in Section \ref{sec:twosol} we show that
two of the parallax solutions can be eliminated by two separate
arguments.  First, they are inconsistant with {\it Kepler} K2 Campaign 9
microlensing observations.  Second, they imply physical effects that
are not observed.
This leaves us with only two solutions, both of which carry the same physical
implications for interpreting the light curve.

\subsubsection{{CMD and {\it Spitzer}-Ground Color-Color Relation}
\label{sec:cmd}}

In order to derive the $VIL$ color-color relations needed to
incorporate the {\it Spitzer} data, we must place the source on a CMD.
We conduct this analysis by first using the OGLE $V/I$ CMD and then
confirm and refine the result using $H$-band data from UKIRT.

The middle panel of Figure~\ref{fig:cmd} shows an instrumental CMD
based on OGLE-IV data.  The clump centroid is at $(V-I,I)_{\rm
  clump,O-IV} = (2.89,16.35)$.  The source is shown at $(V-I,I)_{\rm
  s,O-IV} = (2.57\pm 0.06,21.35\pm 0.01)$, with the color derived
by regression (i.e., independent of model) and the magnitude
obtained from the (final) modeling.  Also shown in this figure are
two points related to the blended light, which are not relevant to
the present discussion but will be important later.  The key point
here is that the source lies 0.32 mag blueward of the clump in the
instrumental OGLE-IV system, which corresponds to 0.30 mag in the
standard Johnson-Cousins system \citep{ogleiv}.

The top panel of Figure~\ref{fig:cmd} shows an $I$ vs. $(I-H)$ CMD,
which is formed by cross-matching OGLE-IV $I$-band to UKIRT $H$-band
aperture photometry.  The magnitude of the clump is the same as in the
middle panel, $I_{\rm clump} = 16.35\pm 0.05$.  To ensure that the $(I-H)$
color of the clump is on the same system as the $(V-I)$ color, we make
a $VIH$ color-color diagram in the lower panel based on cross-matched
stars and then identify the intersection of the resulting track with
the measured $(V-I)$ color to obtain $(I-H)_{\rm clump} = 2.78\pm 0.02$.  Also
shown is the position of the source.  Its magnitude is the same as in
the middle panel.  We determine $(I-H)_s = 2.29\pm 0.03$ from a point
lens model that excludes all data within 3 days of the anomalies.
This permits a proper estimate of the error bars and is appropriate
because the UKIRT data end 3.5 days before the anomaly at peak.
Hence, the source lies 0.49 mag blueward of the clump in $(I-H)$.  To
derive the inferred offset in $(V-I)$ we consult the color-color
relations of \citet{bb88}.  We adopt $(V-I)_{0,\rm clump}=1.08$ from
\citet{bensby13}, which implies $(I-H)_{0,\rm clump}=1.32$ based on
Table~III of \citet{bb88}, and hence $(I-H)_{0,\rm s}=0.83\pm 0.03$.  Then
using Table~II of \citet{bb88}, we obtain $(V-I)_{0,\rm s}=0.75\pm 0.03$,
i.e., 0.33 mag blueward of the clump.  (Note that while we made
specific use of the color of the clump in this calculation, the final
result, i.e., the offset from the clump in $(V-I)$, is basically
independent of the choice of clump color.)\ \ Thus, the results of the
two determinations are consistent.  Although the formal error of the
$I/H$-based determination is smaller than the one derived from OGLE-IV
$V/I$ data, there are more steps.  Hence we assign equal weight to the
two determinations and adopt $(V-I)_0=0.77\pm 0.04$.

To infer the $I-L$ source color from the measured $(V-I)_{\rm
  s,O-IV}=2.57$, we employ the method of \citet{ob161195}.  In brief,
this approach applies the $VIL$ color-color relations of \citet{bb88}
to a $VIL$ cross-matched catalog of giant stars to derive an offset
(including both instrumental zero point and extinction) between the
intrinsic and observed $(I-L)$ color.  Note that in this approach,
explicit account is taken of the fact that the source is a dwarf while
the calibrating stars are giants.  We thereby derive $(I-L)_s =
1.82\pm 0.06$. where here $L$ is the {\it Spitzer} instrumental
magnitude.

From the CMD, we can also derive the angular source size $\theta_*$
(required to determine the Einstein radius $\theta_\e=\theta_*/\rho$).
We adopt a dereddened clump magnitude $I_{0,\rm clump}= 14.35$
\citep{nataf13}.  Using this and the measurements reported above, we
derive $[(V-I),I]_{0,s}=(0.77\pm 0.04,19.35\pm 0.05)$.
Using the $VIK$ color-color
relations of \citet{bb88} and the color/surface-brightness relations
of \citet{kervella04}, this yields (e.g, \citealt{ob03262})
\begin{equation}
\theta_* =(0.455\pm 0.030)\,\muas.
\label{eqn:thetastar}
\end{equation}
where the error is dominated by scatter in the surface brightness
at fixed color (as estimated from the scatter of spectroscopic
color at fixed photometric color, \citealt{bensby13}).
By combining this with $\rho$ measured from the
final model (Section~\ref{sec:kepler}), we
derive\footnote{To avoid ambiguity and possible confusion by
cursory readers, we quote the finally adopted values of the $\theta_*$
and $\theta_\e$ in Equations~(\ref{eqn:thetastar}) and (\ref{eqn:thetaeeval}),
rather than the values derived from the intermediate modeling described thus
far, which differ very slightly.},
\begin{equation}
\theta_\e = {\theta_*\over\rho} = (0.49\pm 0.04)\,\mas
\label{eqn:thetaeeval}
\end{equation}

\subsubsection{{{\it Spitzer} Parallax}
\label{sec:spitzpar}}

Heuristically, space-based microlens parallaxes are derived from the
difference in $(t_0,u_0)$ as seen from observers on Earth and in
space, separated in projection by ${\bf D}_\perp$
\citep{refsdal66,gould94b}.  The vector microlens parallax $\bpi_\e$
is defined \citep{gould92,gould00,gouldhorne}
\begin{equation}
\bpi_\e \equiv {\pi_\rel\over \theta_\e}{\bmu_\rel\over\mu_\rel}.
\end{equation}
Observers separated by ${\bf D}_\perp$ will detect lens-source
separations in the Einstein ring $\Delta {\bf u}\equiv
(\Delta\tau,\Delta\beta) = D_\perp\bpi_\e/\au $, where
\begin{equation}
\Delta\tau \equiv {t_{0,\rm sat} - t_{0,\oplus}\over t_\e},
\qquad
\Delta\beta \equiv u_{0,\rm sat} - u_{0,\oplus}
\label{eqn:dtaudbeta}
\end{equation}
and where the subscripts indicate parameters as determined from the
satellite and ground.  Hence, from a series of such measurements
(which of course are individually sensitive to the magnification and
not to $(t_0,u_0)$ directly), one can infer the vector microlens
parallax
\begin{equation}
\bpi_\e = {\au\over D_\perp}(\Delta\tau,\Delta\beta_{\pm,\pm})
\label{eqn:piederived}
\end{equation}
where $\Delta\beta$ is generally subject to a four-fold degeneracy
\begin{equation}
\Delta\beta_{\pm,\pm} =  \pm |u_{0,\rm sat}| - \pm|u_{0,\oplus}|
\label{eqn:deltabeta}
\end{equation}
due to the fact that in most cases microlensing is sensitive only to
the absolute value of $u_0$ whereas $u_0$ is actually a signed
quantity.

This heuristic picture is somewhat oversimplified because ${\bf
  D}_\perp$ is changing with time, which also means that $t_\e$ is not
identical for the two observers.  Hence, in practice, one fits
directly for $\bpi_\e$, taking account of both the orbital motion of
the satellite and Earth (and hence, automatically, the time variable
${\bf D}_\perp(t)$).  Nevertheless, in most cases (including the
present one), the changes in ${\bf D}_\perp$ are quite small, $|d{\bf
  D}_\perp/dt|\pi_\e t_\e/\au\ll 1$, which means that this simplified
picture yields a good understanding of the information flow.

This qualitative picture can be used to show that the color
constraints play an important role in this event, despite the fact
that the peak is nearly captured in the {\it Spitzer} observations.
As can be seen from Figure~\ref{fig:lc}, in this case {\it Spitzer}
observations begin roughly at peak.  In general, it is quite rare that
{\it Spitzer} observes a full microlensing light curve. This is partly
because the maximum observing window is only 38 days, but mainly
because the events are only uploaded to {\it Spitzer} 3--9 days after
they are recognized as interesting (Figure~1 of \citealt{ob140124}),
which is generally after they are well on their way toward peak.
\citet{yee15} argue that with color constraints, even a fragmentary
lightcurve can give a good parallax measurement. In this case, we have
much more than a fragment, but as we show below, including the color
constraints leads to a much stronger constraint on the parallax
measurement.

If, as in the present case for {\it Spitzer} data,
the peak of the lightcurve is not very well
defined, a free, five parameter $(t_0,u_0,t_\e,f_s,f_b)$, point lens
fit would not constrain these parameters very well.  However, in a
parallax fit, we effectively know $t_\e$, which we approximate here as
identical to the ground, $t_\e=94\,$days.  After applying this
constraint on $t_\e$, fitting the {\it Spitzer} data alone yields $t_{0,\rm
  sat}=7579.5\pm 1.4\,$days and $u_{0,\rm sat}=0.059\pm 0.021$, which
would lead to a parallax error of $\sigma(\pi_\e)\sim 0.021\,\au/D_\perp
\sim 0.016$ and so a fractional parallax error of
$\sigma(\pi_\e)/\pi_\e\sim 40\%$ for the small-parallax solutions.
Note that this would not imply that
the parallax is ``unmeasured'': the fact that the parallax is measured
to be small ($\la 0.06$ and with relatively small errors) would
securely place the lens in or near the bulge, which is significant
information on its location.

However, by including the color-constraint, we can reduce this
uncertainty to $<10\%$, giving a solid measurement of the
parallax. First, one should note that the above fit to the {\it
  Spitzer} light curve yields a {\it Spitzer} source flux of
$f_s=0.22\pm 0.11$ (in the instrumental {\it Spitzer} flux system).
On the one hand, this is perfectly consistent
with the prediction based on the ground solution and the $VIL$
color-color relation $f_{s,Spitz}=0.245\pm 0.015$, which is an
important check on possible systematic errors.  On the other hand, the
errors on the fit value of $f_{s, Spitz}$ are an order of magnitude
larger than the one derived from the $VIL$ relation.  This means that
the color-color relation can significantly constrain the fit. Imposing
this additional constraint, we then find $t_{0,\rm sat}=7579.3\pm
0.8\,$days and $u_{0,\rm sat}=0.0635\pm 0.0029$, substantially
improving the constraints on $u_{0,\rm sat}$. This reduces the
parallax error to about 6\% for the $\Delta\beta_{++}$ and
$\Delta\beta_{--}$ solutions and to about 4\% for the
$\Delta\beta_{+-}$ and $\Delta\beta_{-+}$ solutions.  See Section
\ref{sec:twosol}.

\subsubsection{{Modeling Orbital Motion}
\label{sec:modelorb}}

We now proceed with a simultaneous, 11-geometric-parameter\footnote{Together with, as always, two flux parameters $(f_s,f_b)$ for each observatory, for the source flux and blended flux, respectively.}\hfil\break\noindent
$(t_0,u_0,t_\e,s_0,q,\alpha_0,\rho,\pi_{\e,N},\pi_{\e,E},ds/dt,d\alpha/dt)$
fit to the ground- and space-based data.
The first nine parameters have been described above.  The last
two are a linearized parameterization of orbital motion, with
$\alpha(t)\equiv \alpha_0 + (d\alpha/dt)(t-t_0)$,
$s(t)\equiv s_0 + (ds/dt)(t-t_0)$.

As discussed above, we expect a total of eight solutions: four from
the satellite parallax degeneracy (Equation~(\ref{eqn:deltabeta})) and
two from the two planetary caustics.  However, we found to our
surprise that only one direction of angular orbital motion
was permitted for each of
the four parallax-degenerate solutions, i.e., the source trajectory
could pass through one of the planetary caustics but not the other.
These solutions are given in Table~\ref{tab:ulens_2}.

To understand why only one direction of angular orbital motion is permitted, we
stepped back and performed a series of tests.  In the first
test, we fit for the above 11
parameters but, as in Section \ref{sec:groundmod}, with the data
surrounding the planetary perturbation at $t_p={\rm HJD}^\prime\sim
7500$ removed (specifically $7490<{\rm HJD}^\prime < 7510$).
That is, we removed the information that we had
previously believed was responsible for the measurement of orbital
motion.  Thus, we are testing whether information from the
immediate neighborhood of the planetary caustic is required to predict
the time and position of the planetary caustic crossing.

From the light curve (Figure~\ref{fig:lc}), we can see that the
midpoint of the two caustic crossings of the planetary caustic is $t_p
\sim 7500.375$. From the modeling with the full dataset, we know the
$y$ location of the caustic $\eta_{c,0}$ \citep{han06}. Therefore, if
the orbital motion is constrained by the restricted data set, it
should predict a planetary caustic close to this location. We conduct
this test in a rotated frame for clarity.  For each MCMC sample in
the fit to the restricted data set, we predict the position of the center
of the planetary caustic, first in the unrotated frame according to
\citet{han06},
\begin{equation}
(\xi,\eta_\pm) = \biggl(s - {1\over s},{\pm\sqrt{q}\over s}
\biggl[{1\over \sqrt{1+s^2}} + \sqrt{1-s^2}\biggr]\biggr) .
\label{eqn:xietacaust}
\end{equation}
We then rotate by an angle $\phi=d\alpha/dt(t_p-t_0)$ to obtain
$(\xi^\prime,\eta^\prime_\pm)$, and finally convert this result to
the observational plane
\begin{equation}
(t_x,t_y) = (t_0,0) + (\xi^\prime,\eta^\prime)t_\e
\label{eqn:timeplane}
\end{equation}
The result is shown in Figure~\ref{fig:pred} along with the
``observed'' position of the caustic $(t_x, t_y) = (7500.375, 1.0)$
derived from $(t_p, \eta_{c,0})$.

\begin{figure}
\plotone{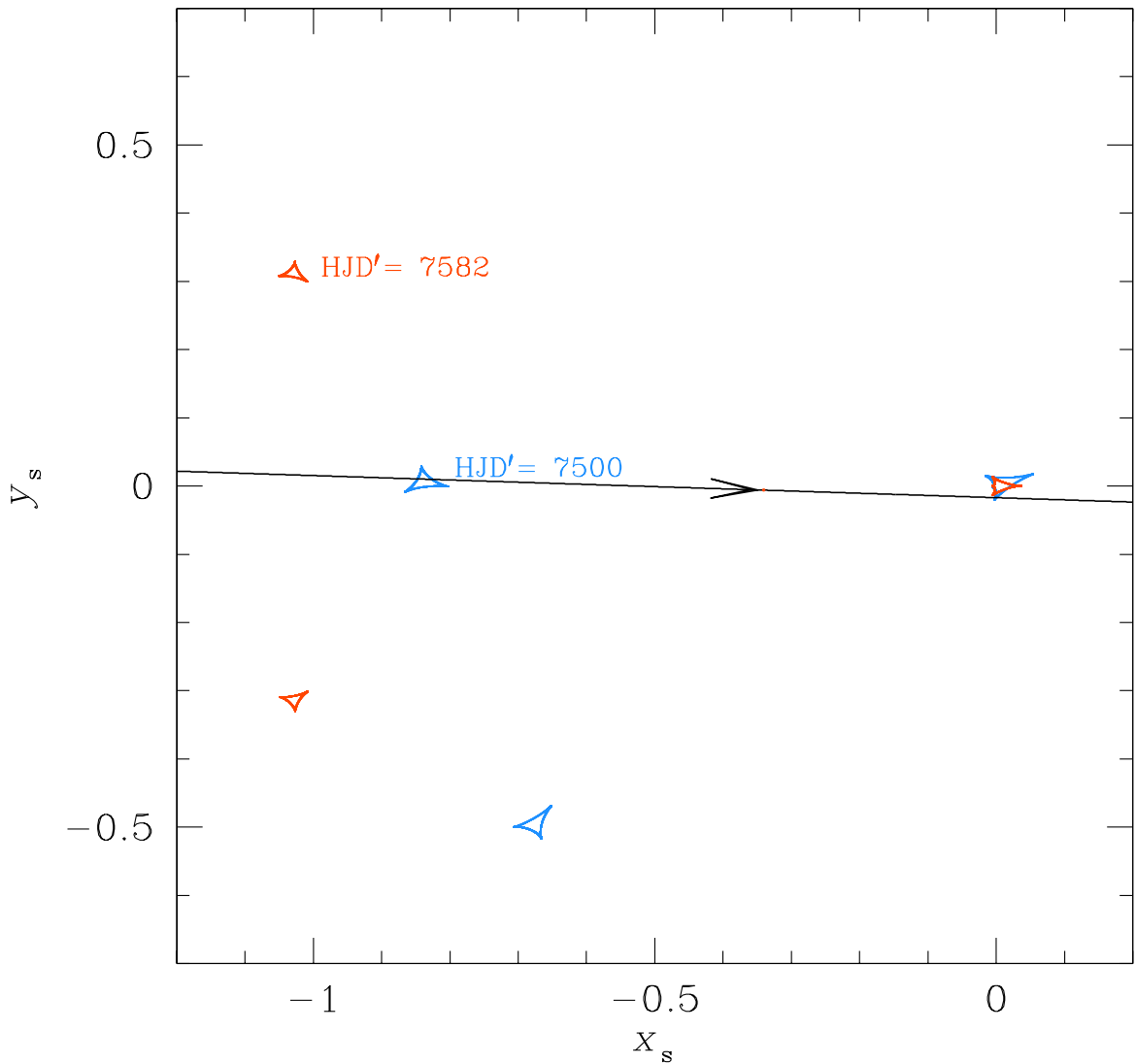}
\caption{Geometry of the source and lens system based on ground-based
data modeled with linearized orbital motion.  The caustic structure is
shown at two epochs, HJD$^\prime=7500$ (blue) and 7582 (red) when,
according to Figure~\ref{fig:lc}, the source has just entered the
planetary caustic and just passed the central caustic, respectively.  A model
that failed to include orbital motion and whose trajectory angle $\alpha$
was determined solely by modeling the source passage over the
central caustic, would miss the (red) caustics.
}
\label{fig:geolin}
\end{figure}

\begin{figure}
\plotone{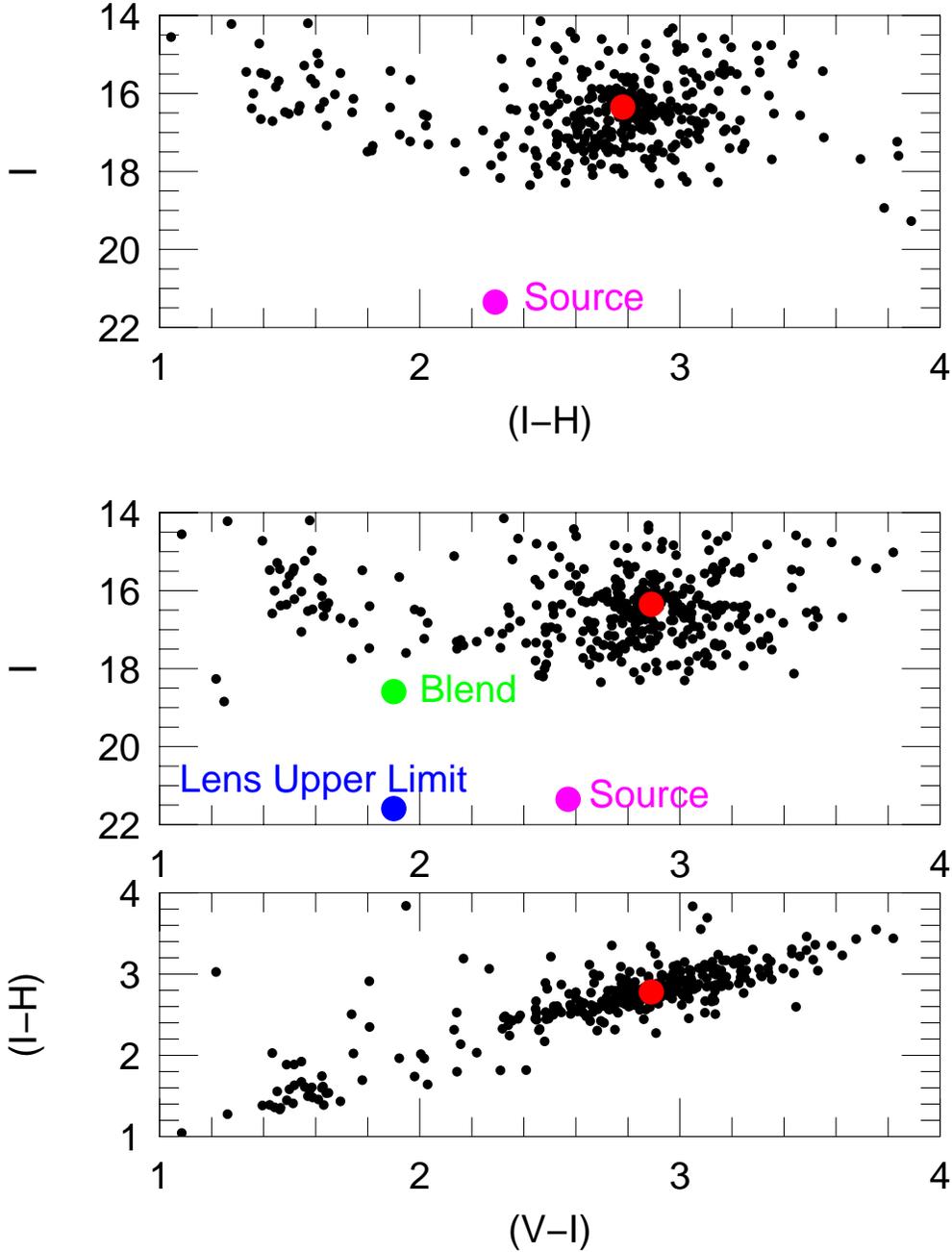} \caption{Instrumental color-magnitude diagrams in
$(I-H,I)$ (top) and $(V-I,I)$ (middle), together with $VIH$
color-color diagram (bottom), which are derived by matching OGLE-IV
instrumental $V$ and $I$ with UKIRT $H$ (aligned to 2MASS).  The
clump centroid is marked in red, while the source is marked in
magenta.  For the $(V-I,I)$ (middle) panel, the blended light is
shown in green. Because the blend is displaced from the source by
$0.5^{\prime\prime}$, only 6\% of its light can be due to the lens.
This flux upper limit shown, in blue (with arbitrary $(V-I)$ color),
restricts the lens mass to $M_L\la 1\,M_\odot$. } \label{fig:cmd}
\end{figure}

There are two main points to note about this figure.  The first is
that the fit to the main light curve, primarily the central-caustic approach,
alone measures the orbital motion
parameters well enough to ``predict'' the position of the caustic to
within a few $\sigma$.  Second, this error bar is quite small, about 2
days in one direction (roughly aligned with time) and 0.5 days in the
transverse direction.  From the inset, which zooms out to the scale of
Figure~\ref{fig:geolin}, one can see that the offset between the
predicted and observed planetary-caustic crossing is tiny compared to
the movement of the caustics that is illustrated in
Figure~\ref{fig:geolin}.

This test demonstrates that the orbital motion can be
determined quite precisely without data from the planetary
caustic, but it does not in itself tell us what part of the
light curve this information is coming from.  In principle,
it could be coming from the cusp approach at the peak of the
light curve or it could be coming from subtle features in the
light curve that lie 10 or more days from the planetary crossing and that
are induced by the planetary caustic itself.  Or, it could be
some combination.  In particular, one suspects that a significant
amount of information must come from the central caustic because
information from the ``extended neighborhood'' of the planetary
caustic would not distinguish between the positive and negative
values of $d\alpha/dt$ that are required for the source
to cross, respectively, the lower and upper planetary caustics
shown in Figure~\ref{fig:geolin}.

Hence, for our second test, we remove all data
$7240<{\rm HJD}^\prime<7567$.  Here we are directly testing what
information is available from the central caustic region.
As shown in Table~\ref{tab:deletedata} (bottom row),
the measurement of the orbital parameters $ds/dt$ and
$d\alpha/dt$ is quite crude compared to either the previous test
or the full data set (first two rows).  Nevertheless, $d\alpha/dt$
is detected at $4\,\sigma$.  Moreover, in order for the direction
of revolution to be in the opposite sense, so that the source
would transit the other caustic in Figure~\ref{fig:geolin},
$d\alpha/dt$ should have the negative of its actual value, i.e.,
$d\alpha/dt\rightarrow +1.42$.  Hence, the value measured after
excluding all data $7240<{\rm HJD}^\prime<7567$ differs from the one
required for opposite revolution by $7.4\,\sigma$.  That is, it is the source
passage by the central caustic that fixes the direction of
the planet's revolution about the host.  Then as can be seen
by comparison of the second and fifth lines of Table~\ref{tab:deletedata},
it is the light curve in the general vicinity of the planetary
caustic that permits precise prediction of the orbital motion
when the data immediately surrounding the caustic are removed.

To further explore the origin of the orbital information,
we show in Table~\ref{tab:deletedata} two additional cases, with
data deleted in the intervals $7495<{\rm HJD}^\prime<7567$ and
$7490<{\rm HJD}^\prime<7567$.  Comparing the last three lines of
Table~\ref{tab:deletedata}, one sees that the lightcurve from more
than 10 days before the crossing contributes greatly to the measument
of transverse motion, and the following 5 days contributes even more.
On the other hand, it is mostly the data after the crossing that
contributes to the measurement of $ds/dt$.

A very important implication of Table~\ref{tab:deletedata}
is that the orbital motion that is predicted based on the the subtle
light curve features away from the planetary-caustic crossings
yield {\it accurate} results.  That is, of the eight hypothetical cases
(2~parameters)$\times$(4~tests) the predictions of orbital motion
are within $1\,\sigma$ of the true value for six cases, and
at $2.0\sigma$ and $2.3\sigma$ in the remaining two.  This provides
evidence that such measurements are believable within their own
errors in other events (i.e., the overwhelming majority) for which
there is no way of confirming the results.

These results have important implications for microlensing observations with
{\it WFIRST} \citep{spergel13}
and, potentially, {\it Euclid} \citep{penny13} because their sun-angle
constraints will very often restrict the light-curve coverage
much more severely compared to those obtained from the ground.

\subsection{{Only The Two ``Large Parallax'' Solutions Are Allowed}
\label{sec:twosol}}

The high precision of the two-parameter orbital motion measurement
motivates us to attempt to model full Keplerian motion.
However, before moving on to this added level of complexity, we first
note that only two of the four solutions permitted by the parallax
degeneracy are allowed.  We present two distinct arguments, the
first based on microlensing and the second based on physical
considerations.

\subsubsection{Degeneracy Breaking From Combined {\it Kepler} and {\it Spitzer} data}
\label{sec:k2}

An important goal for the {\it Spitzer} microlensing program in
2016 was to combine {\it Spitzer} and {\it Kepler} data to break
the 2-fold degeneracy between the two ``large parallax'' solutions
(in which the source passes on opposite sides of the lens as seen
from Earth and {\it Spitzer}) and the two ``small parallax'' solutions
(in which they pass on the same side).  The main
idea for how this would work \citep{prop2015b}
can be understood quite simply in the
approximation that the projected positions of Earth, {\it Kepler},
and {\it Spitzer} are co-linear, i.e.,
${\bf D}_{\perp,K2}=g{\bf D}_{\perp,Spitz}$, near the peak of the event,
where $g$ is roughly constant over short time periods.  Since
all three bodies are in or very near the ecliptic, this approximation
would be almost perfect for microlensing events in the ecliptic
and is still quite good for the K2 field, which lies a few degrees
from the ecliptic ($4^\circ$ for this event).  In this case (as one
may easily graph),
\begin{equation}
u_{0,k2} = g u_{0,Spitz} + (1-g)u_{0,\oplus}
\label{eqn:u0interp}
\end{equation}
and therefore
\begin{equation}
{|u_{0,k2,\pm,\pm}|\over|u_{0,k2,\pm,\mp}|
} = {|g |u_{0,Spitz}| - (1-g)|u_{0,\oplus}||\over
g |u_{0,Spitz}| + (1-g)|u_{0,\oplus}|
}.
\label{eqn:u0rat}
\end{equation}
This formula can be more directly comprehended in the approximation
(quite robust in the present case) that $A_\max \simeq |u_0|^{-1}$.
Then,
\begin{equation}
{A_{\max,k2,\pm,\pm}\over A_{\max,k2,\pm,\mp}
} = {g A_{\max,\oplus} + (1-g)A_{\max,Spitz}\over
|g A_{\max,\oplus} - (1-g)A_{\max,Spitz}|
}.
\label{eqn:arat}
\end{equation}
That is, the small parallax solutions ($(++)$ and $(--)$) will always
yield higher peak magnifications for {\it Kepler}, unless either
$u_{0,Spitz}=0$ or $u_{0,\oplus}=0$, in which case the ``large'' and
``small'' parallax solutions are equal anyway \citep{gould12}.
In the present case, $g\simeq 5/8$ and
$A_{\max,\oplus}/A_{\max,Spitz}\simeq 4$.  Hence, Equation~(\ref{eqn:arat})
predicts $A_{\max,k2,\pm,\pm}/ A_{\max,k2,\pm,\mp}\simeq 23/17\sim 1.35$.
This is quite close to the more precise value from a rigorous numerical
model, which is illustrated in the main panel of Figure~\ref{fig:k2_lc}.
(The offset between the two curves is less that the naive 0.33 mag
because the curves are aligned to the ground-based light curve,
which is heavily blended.)

Unfortunately, as also shown in Figure~\ref{fig:k2_lc},
there are no {\it Kepler}
data over peak because the K2 C9 campaign ended nine days earlier.
Moreover, as shown in the lower right panel, the large and small
parallax models predict almost identical {\it Kepler} light curves
in the region of the approach to the peak where there are data.

However, as shown in the lower left panel, the two models predict
dramatically different light curves at the time of the first
(planetary) caustic.  While we have shown only one realization
from the MCMC chain (the one with best $\chi^2$), we find that
these predicted differences are extremely robust for all solutions
with reasonable $\chi^2$.  This robustness can be understood from
Figure~\ref{fig:k2_caust},
which shows the source trajectories as seen from Earth
and {\it Kepler} in the neighborhood of the planetary caustic
for each of the four solutions.

The first point to note about these four panels is that while
the caustic is not in the same place or same orientation in the Einstein ring
(because the geometric parameters of these solutions are not exactly
the same), the path of the source relative to the caustic as
seen from Earth is extremely similar.  This is simply because this
path is directly constrained by the ground-based data that are shown in
the lower left panel of Figures~\ref{fig:lc} and \ref{fig:k2_lc},
and in the further zoom of this region shown in
Figure~\ref{fig:lc_zoom}.
Second, if we look at
each sub-figure separately, we see that the vector offset between
the source positions seen from Earth and {\it Kepler} barely changes.
This reflects that fact that over this short, three-day interval,
${\bf D}_{\perp,k2}(N,E)\simeq (-0.05,-0.13)\au$ is nearly constant.

The primary difference between the two upper panels is that for
$(-,+)$, the source passes above the Earth trajectory as seen from
{\it Kepler}, whereas for $(-,-)$ it passes below.  Since the
caustic is narrower toward the top, the source has already exited
when the observations begin (magenta circle) for the $(-,+)$ case.
Contrariwise, since the caustic widens toward its base, the earliest
{\it Kepler} data are still inside the caustic for the $(-,-)$ case.
Moreover, since the base of the caustic is ``stronger'' than the
middle, the spike from the caustic exit is more pronounced than
it is from Earth.

What is the reason for this opposite behavior?  In both cases
$u_{0,\oplus}<0$, meaning that, by definition
(Figure~4 from \citealt{gould04}), the source passes the lens on its
left.  Then, in the two cases, the source passes the lens as seen
from {\it Kepler} on its left and right, respectively, implying that the source
is displaced from Earth to opposite sides from the direction
of motion.

A secondary effect is that {\it Kepler} is slightly leading
Earth for $(-,+)$ and slightly trailing for $(-,-)$, which also
contributes to the fact that the caustic exit does not occur ``in time''
for the start of the {\it Kepler} observations in the first case, but
does in the second.  These effects can be derived from the general
formula for the {\it Kepler}-Earth offset relative to the
direction of the source trajectory (following the formalism
of \citealt{gould04}),
\begin{equation}
(\Delta\tau,\Delta\beta)={(-\bpi_\e\cdot{\bf D}_\perp,-\bpi_\e\times{\bf D}_\perp)
\over {\rm AU}}
\label{eqn:deltataubeta}
\end{equation}
However, since this is a secondary effect, we do not present details
here.

Because microlensing fields are very crowded compared to the original
{\it Kepler} field for which the camera was designed, there are
usually several stars within a {\it Kepler} pixel that are much
brighter than the microlensed source, even when it is magnified.
This, combined with the not-quite regular 6.5-hour drift cycle
in K2 pointing means that standard photometry routines cannot
be applied to K2 microlensing data.  Here we employ the algorithm
of \citet{huang15} and \citet{soares17}, which was originally
developed for other crowded-field applications and then further developed
by  \cite{zhu17b} for microlensing.
We refer the reader to these papers for details of the method.
However, an important point to emphasize is that the method requires
detrending, which can partly or wholly remove long term
features.  Hence, for example, it could be problematic to apply
it to the long, slow rise predicted for {\it Kepler} in the weeks
before Campaign 9 ended on $\sim 7572.4$.  Fortunately, as we have
discussed, there is no possibility of distinguishing the models from
this part of the light curve in any case.

Instead, we are only interested to determine whether there is a
sharp ``spike'' in the K2 light curve shortly after the observations
begin.  Since detrending must be done separately on the two sections
of K2 lightcurves (before and after the hiatus to download data beginning
at ${\rm JD}^\prime = 7527.4$), we restrict the detrending to before this
date.  We also restrict ${\rm JD}^\prime > 7502.5$ to avoid the region
of the light curve that could conceivably be impacted by the possible
``spike''.   In this interval, all microlensing models agree that
the K2 light curve is essentially flat, so that it can be ``modeled''
as a constant.  Note that since no
microlensing model is required to construct this light curve, the
specific modifications introduced by \citet{zhu17b} are not
actually required for these reductions.

Figure~\ref{fig:k2obs} shows the detrended K2 light curve together
with the four otherwise-degenerate microlensing models.
The ``small parallax'' models [$(+,+)$ and $(-,-)$] each predict a sharp
spike due to a caustic exit shortly after the onset of the K2 C9
campaign, whereas the ``large parallax'' models do not.  In order to
transform the model magnification curves to predicted K2 light curve,
one must determine the {\it Kepler} magnitude $K_p$ of the source.
We first determine the calibrated $I$ and $V$ magnitudes of the source
by aligning the OGLE-IV source magnitudes to the calibrated OGLE-III
\citep{ogleiii1,ogleiii2}
system $(V-I,I)=(2.44,21.47)$.  We then incorporate extinction parameters
from \citet{nataf13} and apply the transformations given in \citet{zhu17b}
to find $K_p = 23.03\pm 0.10$.  That is, the model curves have a systematic
scaling error of 10\% flux.  From Figure~\ref{fig:k2obs} it is clear that
the K2 data are inconsistent with the ``small parallax'' models,
even allowing for this 10\% error in the scale of the models.

Of course, the entire argument given in this section depends
on the model being correct within its stated errors.  We mentioned
above that all models that are consistent with the data yield
extremely similar light curves in the neighborhood of the caustic,
which, as we emphasized, is not at all surprising given that the
source trajectory relative to the caustic is directly determined
by the data.

In principle, there remains the question of whether
the data themselves have systematics that would incorrectly constrain
the model to this particular geometry.
Figure~\ref{fig:lc_zoom} shows that this is unlikely
because on the principal defining feature, the caustic entrance,
the data (particularly KMTA) have errors ($\sim 0.05\,$mag)
that are small compared to the compared to the caustic entrance
``jump'' ($\sim 0.25\,$mag).  Nevertheless, given the apparent
importance of these data to the final result, we conduct
four tests that alter the data around the planetary caustic
$(7490.5< {\rm HJD}^\prime < 7510.5)$: 1) remove MOA data, 2) remove KMTA
data, 3) bin both MOA and KMTA data, 4) remove both MOA and KMTA data.
We find that the fit parameters change by $\ll 1\,\sigma$
when MOA data are removed and by $< 1\,\sigma$ when KMTA data are removed.
These first two tests essentially rule out that the result can be
strongly influenced by systematics, since it is extremely unlikely
that the systematics would be the same at observatories separated
by thousands of km.  Even when both data sets are removed, the
results change by $\la 1\,\sigma$.  Binning the data also
affects the results by $< 1\,\sigma$.

It may seem somewhat surprising
that even elimination of both MOA and KMTA do not prevent
the model from precisely locating the planetary caustic given that
these data sets alone probe the caustic entrance.  However, the
size, shape and orientation of the caustic are precisely specified by
the parameters $[s(t_c),q,\alpha(t_c)]$, which are well determined from the
overall model.  Here, $t_c = 7500$ is the time of the planetary caustic.
Given this, the facts that the OGLE and KMTC data probe the internal
height of the caustic at two epochs, while the KMTS (and also KMTC) data
define the post-caustic cusp approach, constrain the position of
the caustic quite well.

We conclude that the analysis
given in this section is robust against both statistical and
systematic errors.

\subsubsection{Degeneracy Breaking From Physical Constraints}
\label{sec:degenphys}

Next we give a completely independent argument that essentially
rules out the ``small parallax'' solutions.  Almost by definition
(Equation~(\ref{eqn:deltabeta})),
$|\Delta\beta_{\pm,\pm}|<|\Delta\beta_{\pm,\mp}|$, and therefore
$|\pi_{\e,{\pm,\pm}}|<|\pi_{\e,{\pm,\mp}}|$.  Hence, the masses
$M=\theta_\e/\kappa\pi_\e$ for the $(+,+)$ and $(-,-)$ solutions are
higher than the other two solutions.  In the present case, they are
higher by a factor $\sim 1.75$, which would put the primary at
$M=1.60\pm 0.16\,M_\odot$.  See Table~\ref{tab:phys_2}.

There are two arguments against such a heavy lens.  First, it would
give off too much light.  Second, given its almost certain location in
the Galactic bulge, its main-sequence lifetime ($\la 2\,$Gyr) would
be too short given the typical range of ages of bulge stars.  Both of
these arguments implicitly assume that the host is not a neutron star,
which we consider to be extremely unlikely given the complete absence
of easily detectable massive companions like OGLE-2016-BLG-1190Lb
at a few AU around pulsars.  See Figure~1 of \citet{martin16} and
note that while the black points have similar masses to OGLE-2016-BLG-1190Lb,
they have semi-major axes $a\sim R_\odot$ and hence are likely to be
stripped stars rather than planets and, in any case, not at $a\sim \au$.

The blended light shown in Figure~\ref{fig:cmd} $[(V-I),I]_{\rm
  b,O-IV}=(1.90,18.59)$ lies just 2.3 mag below the clump and is about
1 mag bluer than the clump.  Hence, it would not be at all
inconsistent with a roughly $1.6\,M_\odot$ star at the distance of the
Galactic bulge.  However, although the source star is intrinsically
faint, its position can be determined with high precision when it is
highly magnified, from which we determine that the light centroid of
the blend is offset from the position of the blend by
$0.5^{\prime\prime}$.  By examining the best-seeing OGLE-IV baseline
images, we find that the total light at the position of the source
must be less than 13\% of the blended light.  Since (from the fit)
$f_s/f_b = 0.07$, the light due to the lens must be at least 3 mag
fainter than blend, and hence 5.3 mag fainter than the clump.  This is
clearly inconsistent with a star in the bulge that is significantly
more massive than the Sun.

In addition, the lens-source relative parallax for the $(+,+)$ and $(-,-)$
solutions is $\pi_\rel\simeq 20\,\muas$.  The source-lens relative distance
is given by $D_{LS}\equiv D_S-D_L=\pi_\rel D_L D_S/\au\simeq 1.2\,\kpc$.  This
small separation, combined with the fact that the source color and magnitude
are quite consistent with it being a bulge star, imply that the lens
is heavily favored to lie in the bulge.  The bulge is generally thought
to be an old population.  If this were strictly true, it would rule
out such massive bulge lenses.  However, \citet{bensby17} find that
their sample of microlensed bulge dwarfs and subgiants has a few percent
of stars with ages less than 2 Gyr.  Thus, while this second argument
against the $(+,+)$ and $(-,-)$ solutions is less compelling than the
first argument, it does tend to confirm it.

\subsection{{Full Keplerian Orbital Solution}
\label{sec:kepler}}

To investigate full Kepler solutions, we add two new parameters and
transform the meaning of two previous parameters to
obtain\hfil\break\noindent
$(t_0,u_0,t_\e,s,q,\alpha,\rho,\pi_{\e,N},\pi_{\e,E},ds_{\bot}/dt,d\alpha/dt,
s_{\parallel},ds_\parallel/dt)$; we also specify the reference
time\footnote{As a matter of convenience, we have set $t_{\rm
binary}$ (the zero point of the orbital solution) near $t_0$.
However, in contrast to $t_0$ it is held fixed and so does not vary
along the chain.} $t_{\rm binary} = 7582.16$.  Here, the two triples
$(s,0,s_{\parallel})$ and
$[ds_{\bot}/dt,s(d\alpha/dt),ds_\parallel/dt]$ are, respectively,
the instantaneous planet-star separation and relative velocity, in
the coordinate frame defined by the planet-star axis on the sky, the
direction within the plane of the sky that is perpendicular to this,
and the direction into the plane of the sky. The units are Einstein
radii and Einstein radii per year, so that to convert to physical
units one should multiply by $D_L\theta_\e$.  Of course, if these
parameters are specified, together with the total mass of the
system, one can determine the full orbit and hence the Kepler
parameters.

\citet{ob09020} discusses the transformations from microlensing
parameters to Kepler parameters in detail.  For each MCMC sample,
one determines $\theta_*$ from the value of $f_s$ and the
model-independent color.  Then from the value of $\rho$, one
obtains $\theta_\e=\theta_*/\rho$.  Then from the value of the
parallax $\bpi_\e$, one determines $M=\theta_\e/\kappa\pi_\e$ and
$\pi_\rel = \theta_\e\pi_\e$.  In order convert the position and
velocity parameters into physical separations and velocities, one
still needs $D_L=\au/(\pi_\rel + \pi_s)$ where $\pi_s$ is the source
parallax.  For this we adopt $D_s=8.7\,\kpc$, as discussed below in
Section \ref{sec:physchar}.  We report the microlens parameters for
the two remaining solutions in Table~\ref{tab:ulens_4} and show the
MCMC sampling of parameters for one of these solutions in
Figure~\ref{fig:mptri}.
We also show the transformation of this sampling to the key Kepler
parameters in Figure~\ref{fig:mpphys}.

Figures~\ref{fig:mptri} and \ref{fig:mpphys} show that while the
microlens parameters exhibit well-behaved, relatively compact
distributions, the Kepler parameters follow complex one-dimensional
structures. This is probably due primarily to the fact that one of
the two new parameters $s_{\parallel}$  is relatively well
constrained, while the other $ds_\parallel/dt$ is fairly poorly
constrained.  As a result, some of the Kepler parameters are also
poorly constrained.  In particular, unfortunately, it is not
possible to strongly constrain the eccentricity, which would have
been a first for microlensing.
Nevertheless, we note that OGLE-2016-BLG-1190Lb has the best
constrained orbit of any microlensing planet yet detected.

Figure~\ref{fig:geo} shows the geometry of the source and lens system
together.  The dashed black line shows the planet's orbit, with its
position at the times of the two caustic crossings shown by orange
dots.  The caustic structure at the first of these epochs is shown in
blue and at the second in red.  The trajectories of the source
position through the Einstein ring as seen from {\it Spitzer} and
Earth are shown in green and black respectively.  The curvature in
these two trajectories reflects the orbital motion of these two
observatories about the Sun.

Table~\ref{tab:compchi2} summarizes the evolution of models
developed in this paper.  The penultimate column gives the change in
$\chi^2$ relative to the previous model, for the set of
observatories specified in the final column.  The ``standard model''
(no higher order effects) was not presented here in the interest of
space and is shown for reference in order to emphasize the enormous
improvement in $\chi^2$ by introducing two orbital parameters. We
introduced parallax at the same time that we included {\it Sptizer}
data.  The fact that $\chi^2_{\rm gr}$ decreases by
$\Delta\chi^2_{\rm gr}=10.6$ means that the ground-based data
corroborate the much more precise {\it Spitzer} parallax measurement
at this level.  The final $\Delta\chi^2=9.6$ improvement when full
orbital motion is included may seem rather modest, and within
conventional reasoning on this subject it might even be questioned
whether these two extra parameters have been ``detected''. However,
as pointed out by \citet{ob150479}, since binaries are known a
priori to have Keplerian orbits, the only justification required for
introducing additional parameters is that they can be measured more
precisely than they are known a priori.

\section{Physical Characteristics}
\label{sec:physchar}

The physical parameters derived from the full orbital solutions are
given in Table~\ref{tab:phys_4}.
As discussed in Section~\ref{sec:twosol},
the blended light in the CMD (Figure~\ref{fig:cmd}) provides
a constraint on the lens mass, independent of the modeling.
The light superposed on the source (including the lens and possibly
other stars) is at least 5.3 mag below the clump, which excludes
lenses that are significantly more massive than the Sun.
As discussed there, this essentially rules out the $(+,+)$ and $(-,-)$
solutions.  For the $(+,-)$ and $(-,+)$ solutions, since very few of the MCMC
samples exceed this mass limit, we do not bother to impose this constraint.
However,
we note that given the mass measurements reported in Table~\ref{tab:phys_4},
together with the faintness of the source shown in Figure~\ref{fig:cmd},
the lens should be easily detectable in high resolution images,
a point to which we will return below.

The main focus of interest is the mass of the companion and the distance
of the system,
\begin{equation}
M_2 = 13.4\pm 0.9\,M_J;
\qquad
D_L = 6.8\pm 0.1\,\kpc.
\label{eqn:massdist}
\end{equation}
The mass is at the edge of the conventional planetary range, i.e.,
a massive super-Jupiter, very close to
the deuterium limit that conventionally separates ``brown dwarfs'' from
``planets''.

As discussed by \citet{21event}, what is actually measured precisely in
microlensing events is $\pi_\rel = \theta_\e\pi_\e$ rather
than $D_L = \au(\pi_\rel + \pi_s)^{-1}$ because the precise value of
$D_s$ (and so $\pi_s=\au/D_S$) is not known.  The uncertainty in $\pi_s$
becomes particularly important for lenses in and near the bulge because
it dominates the error in $D_L$.  For example, in the present case
$\pi_\rel = 33\pm 3\,\muas$, whereas a 7\% error in $D_S$ would
yield an error in $\pi_s$ almost three times larger than this.

For this reason, \citet{21event}, introduced a deterministic
tranformation of $\pi_\rel$ that could be used in statistical studies
of {\it Spitzer} microlensing events, $D_{8.3}$ which can be evaluted
in the present case\footnote{Note that for
lenses in or near the bulge, the fractional error in $D_{8.3}$ is,
by the chain rule, much smaller than the fractional error in $\pi_\rel$.}
\begin{equation}
D_{8.3} \equiv {\kpc\over \pi_\rel/\mas + 1/8.3}= 6.54\pm 0.13\,\kpc.
\label{eqn:d_8.3}
\end{equation}
This formulation has the advantage that it is symmetric with respect
to nearby and distant lenses.  That is, for lenses near the Sun,
$D_L\rightarrow D_{8.3}$ while for lenses near the source,
$D_{LS}\equiv D_S - D_L \rightarrow 8.3\,\kpc - D_{8.3}$.  Hence,
because $D_{8.3}$ is more precisely determined than $D_L$,
particularly for lenses in and near the bulge, it is more suitable
for statistical studies and for comparison of different events.  To
enable such comparisons, we always use 8.3 (kpc) in the denominator
of Equation~(\ref{eqn:d_8.3}) rather than the best estimate of the
source distance.  However, there is very little cost to this
approximation in terms of its impact on the quantity of physical
interest for near bulge lenses, $D_{LS}$, which is approximated by
$D_{LS,8.3}\equiv 8.3\,\kpc - D_{8.3}$  For example, in the present
case
\begin{equation}
{D_{LS,8.7} - D_{LS,8.3}\over D_{LS,8.7}} \rightarrow {1.91\,\kpc - 1.76\,\kpc
\over 1.91\,\kpc} = 8\% .
\label{eqn:dlstest}
\end{equation}

We note that the distance shown in Equation~(\ref{eqn:massdist})
adopts the same uncertainty as Equation~(\ref{eqn:d_8.3}) and neglects
%
%
the uncertainty due to the
source distance. In this case, we have derived the above lens distance
by fixing the source distance at $8.7\,\kpc$, i.e., about $0.9\,\kpc$
behind the center of the ``bulge'' (really, ``bar'') toward this line
of sight.  We have chosen this distance as typical because, as we
argue just below, the lens and source are both likely to lie in the
Galactic bar.

Both the measured values of $\pi_\rel$ and $\mu_\rel$ indicate that
the most probable configuration is that the lens and source are
roughly equally displaced from the center of the bar. The measured
value of $\pi_\rel$ leads to an inferred lens-source separation
$D_{LS}=D_LD_S\pi_\rel/\au = 2.1\pm 0.2\,\kpc$. This is consistent
with the lens being in the bar, but it does not by itself argue
strongly for such an interpretation.  The lens could also lie $\sim
2\,\kpc$ in the front of the bar, in the Galactic disk.  However, in
this case, we would expect the lens-source relative proper motion to
be substantially greater than the one that is measured:
$\mu=1.9\,\masyr$.  This small value of the relative proper motion is
more consistent with a source and lens drawn from kinematically
related populations of stars.  Thus, we tentatively conclude that this
very massive super-Jupiter companion to a G dwarf is in the
bar/bulge and adopt a source distance $D_s = 8.7$ kpc.

However, at this stage, a disk lens cannot be ruled out.  In particular,
if the source happened to be moving at $\sim 5\,\masyr$ in the direction
of Galactic rotation, then the inferred motion for the lens
(based on the observed $\bmu = \mu(\bpi_\e/\pi_\e$)) would be quite
consistent with disk kinematics.

Because the lens and source are almost certainly detectable in high
resolution imaging (whether space-based or ground-based), this question
about the lens kinematics can be resolved within a few years.
Under most circumstances, such followup observations require
that the source-lens {\it relative} motion be measured (which could
take quite some time given their low relative proper motion). In the
present case, the relative motion is small meaning that the source and
lens will remain unresolved for many years, but we can still measure
their {\it joint} proper motion relative to the stellar
background.
Two
measurements separated by two years should be sufficient to determine
this joint motion relative to the frame of field stars.  The absolute
motion of this frame can then be determined from {\it Gaia} data.
Since the lens and source are hardly moving relative to one another,
this should indicate the kinematics of the lens star.  Finally,
one can make a final correction from the observed, joint proper motion
to the lens proper motion based on the well measured magnitude
and direction of the lens-source relative proper motion
(from the microlens solution) together with the ratio of the
source flux (also from the microlens solution) to the total flux
(from the high-resolution data themselves).

To aid in such future observations we report the $(I-H)_s$ color
of the source, as described in Section~\ref{sec:cmd} and also the
$H_s$ magnitude of the best-fit model
\begin{equation}
(I_{\rm ogle} - H_{\rm 2mass})_s = 2.29\pm 0.03;
\qquad
H_s = 19.06\pm 0.04
\label{eqn:imh}
\end{equation}

\section{{Assignation to {\it Spitzer} Parallax Sample}
\label{sec:assign}}

The Galactic distribution of planets experiment must be carried out
strictly in accord with the protocols specified by \citet{yee15},
which were designed to maintain an unbiased sample. Because
OGLE-2016-BLG-1190 was selected for {\it Spitzer} observations on the
basis of the planetary anomaly observed at $\sim 7500$, it would
naively appear that this planet cannot be included in a sample that
must be unbiased to planets. However, \citet{yee15} anticipated just
such a situation (in which the planetary anomaly occurs early in the
light curve) and so laid out specific criteria under which such
planets may be included while still maintaining the objectivity of the
sample.

\citet{yee15} specify several ways an event may be selected for {\em
  Spitzer} observations, two of which are relevant here. The first is
following pre-specified ``objective'' criteria under which planets
found during {\it any part of the event} can be included in the
sample. If an event meets certain purely objective criteria, then it
{\it must} be observed by {\it Spitzer} at a specified minimum cadence
(usually once per day). Observations can only be stopped according to
specified conditions. Since there is no human element involved in event
selections of this type, any planets and planet sensitivity automatically enter the
unbiased sample. However, {\it Spitzer} time is extremely
limited and ``objective'' selection places a large and rigidly
enforced burden on this time, so the ``objective'' criteria must be
set conservatively.

Therefore \citet{yee15} also specify
the possibility of ``subjective'' selection, which can be made for any
reason deemed appropriate by the team.  In this case, only planets
(and planet sensitivity) from data not available to the team at the
time of their decision can be included in the sample.  Specifically,
if a planet (or a simulated planet used to evaluate planet
sensitivity) gives rise to a deviation from a point-lens model in such
available data that exceeds $\Delta\chi^2=10$, then it must be
excluded.  This effectively removes not only ``known'' planets but
also ``unconsciously suspected'' planets.  The cadence and conditions
for stopping subjectively alerted events must be specified at the time
they are publicly announced.

Although ``subjective'' selection has the obvious disadvantage that
only planets discovered after the selection can be included in the
sample, there are several advantages to this type of selection. Many
of these advantages derive from the fact that {\it Spitzer} targets
can only be uploaded to the spacecraft once per week and must be
finalized $\ga 3$ days before observations begin. First, an event
may never become objective and yet still be a good candidate. For
example, if it is short timescale and peaks in the center of the {\it
  Spitzer} observing ``week,'' it may still be too faint to meet the
``objective'' criteria when the decision to observe must be made and
again may be too faint the following week. Second, the team may select
an event ``subjectively'' a week or two before it meets the
``objective'' criteria. In that case, {\it Spitzer} observations start
a week or two earlier, improving the measurement of the
parallax. Likewise, the team may specify a higher {\it Spitzer}
cadence for a ``subjectively'' selected event, resulting in more
observations and a better parallax.

\citet{yee15} also specify how to classify an event that may be
selected multiple ways, such as a ``subjective'' event that later
meets the ``objective'' criteria. From the perspective of measuring a
planet frequency, ``objectively'' selected events are clearly better
because then planets from the entire light curve can be
included. However, from the perspective of measuring the Galactic
distribution of planets, an event is worthless if the parallax is not
measured. Thus, \citet{yee15} state that the ``objective''
classification takes precedence as long as the parallax is
``adequately'' measured from the subset of the data that would have
been taken in response to an ``objective'' selection, i.e., after
removing any data taken before the ``objective'' observations would
start and thinning the data to the ``objective'' cadence. If the
parallax is not ``adequately'' measured from this subset of the data,
then the event reverts to being ``subjectively'' selected and,
consequently, all planets detectable in data available prior to the
selection date are excluded. However, if an event is determined to be
``objective'' based on these criteria, all {\it Spitzer} data can be
used to {\it characterize} the lens.

\citet{yee15} do not specify what precision is required for an
``adequate'' parallax, but this was discussed in \citet{zhu17}. They
established a condition
\begin{equation}
\sigma(D_{8.3})< 1.4\,\kpc. \label{eqn:D8.3}
\end{equation}
We note that \citet{21event} introduced $D_{8.3}$ because $\pi_\rel$
is more reliably measured than $D_L$ and because this quantity gives
symmetric information on the distance between the observer and lens
for nearby lenses and between the source and lens for distant lenses.

OGLE-2016-BLG-1190 was clearly initially selected ``subjectively,''
and under those criteria, the planet could not be included. In Section
\ref{sec:yee_proc}, we show that this event actually met the
``objective'' criteria. Then, in Section \ref{sec:zhu_proc}, we
examine whether or not the data taken in response to the ``objective''
selection meet the \citet{zhu17} criterion (Equation \ref{eqn:D8.3})
for an adequately measured parallax.

\subsection{Yee et al. (2015) Protocols: Is the Event Objective?}
\label{sec:yee_proc}

As discussed in Section~\ref{sec:spitzerobs}, OGLE-2016-BLG-1190 was
initially selected on 1 July 2016 solely because the pre-alert light
curve contained a candidate planetary anomaly.  A check at the time of
the initial upload to {\it Spitzer} (4 July 2016) showed that it did
not meet objective criteria.  As we now show, however, at the upload
the following week on Monday 11 July 2016, the event did meet the
objective criteria for rising events (B1--5) of \citet{yee15}.  We
note that while the {\it Spitzer} team does make some effort to
determine which already-selected events have become ``objective'',
this is not carried out uniformly, nor is it required by the
\citet{yee15} protocols.  No such effort was made for
OGLE-2016-BLG-1190, so we are doing it here for the first time.

There are two sets of ``objective'' criteria: ``falling'' criteria
that take into account the model fit and ``rising'' criteria based
almost entirely on model-independent observables. The reason for the
distinction is that the model parameters generally remain uncertain
until after an event has peaked.  For rising events, there is only one
model-dependent criterion, i.e., that according to the best fit model,
the event has not yet peaked or is less than 2 days past peak.  Here
the idea is that either the event has already peaked, in which case
there is a good model for when that peak occurred, or it has not
peaked in which case no plausible model will say that it peaked more
than 2 days previously. On 11 July, OGLE-2016-BLG-1190 was clearly
pre-peak and therefore should be judged under the ``rising'' criteria.

For an event to meet the ``rising'' criteria, it must be in a
relatively high cadence OGLE or KMTNet field, which as described in
Section~\ref{sec:groundobs} is clearly satisfied.  Then, there are
three flux criteria that must be satisfied. First $I_{\rm now}<17.5$,
second $I_{\rm now}<I_{\rm base}-0.3$, and third $I_{\rm now}<16.3+
0.93 A_I$, where $A_I$ is the $I$-band extinction toward this field.
Since $A_I>1.3$ and $I_{\rm base}<17.8$, the operative condition is
$I_{\rm now}<17.5$.

To assess whether or not OGLE-2016-BLG-1190 met these flux criteria,
we must be careful to make use of the data only as they were available
to the team at the time of the final decision, UT 13:30 11 July
(HJD$^\prime =7581.06$).  This means not only truncating the data at
that date, but also using the versions of the data sets that were
available and verifying that these were in fact available.  MOA data
are accessible in real-time while the OGLE data are generally delayed
by of order 12 hours. We check that the last such OGLE data point was
posted at HJD$^\prime =7580.04$.  In addition, the magnified-source
flux derived from OGLE data that were available online at that time
are fainter than the re-reduced data used in the analysis (Section
\ref{sec:analysis}) by an average of $\sim 0.23$ mag.  This is because
the online data were obtained using a catalog star whose position was
displaced by $0.5^{\prime\prime}$ from the true source\footnote{One
  expects for, e.g., a Gaussian profile, a flux loss corresponding to
  $\Delta I=(\log 32)(a/{\rm FWHM})^2\simeq 1.5(a/{\rm FWHM})^2$ where
  $a=0.5^{\prime\prime}$ is the offset.  That is $\Delta I\simeq 0.25$
  for ``typical good seeing'' ${\rm FWHM}\sim 1.2^{\prime\prime}$.},
as noted in Section~\ref{sec:twosol}.  Since the OGLE scale is used by
the {\it Spitzer} team for the flux determinations, we must use the
online OGLE data (rather than the re-reduced data) to assess whether
or not OGLE-2016-BLG-1190 met the flux criteria.

We fit the online OGLE and MOA data HJD$^\prime <7581.06$ to a
point-lens model (note that no OGLE data were taken between 7579.762
and 7581.719).  This fit shows that the event reached $I=17.5$ on the
online-OGLE scale about 1.65 days before the upload deadline.  This
means that the MOA data points taken beginning 1.3 days before this
deadline, which could be aligned to the OGLE scale, were already above
the threshold, leaving no doubt that the event became objective well
in advance of the upload time.

\subsection{Zhu et al. (2017) Protocols: Is the Parallax Measured?}
\label{sec:zhu_proc}

The parallax is clearly measured from the analysis of the full dataset
(Section \ref{sec:analysis}). We now must determine whether or not the
parallax is ``objectively'' measured and meets the \citet{zhu17}
criterion in Equation~(\ref{eqn:D8.3}). First, we can only include the
{\it Spitzer} data starting at 7585.98, the date objective {\it
  Spitzer} observations would have begun. Second, we must determine
whether it would have met the \citet{zhu17} criterion {\it if it had
  been a point-lens event}.  This is because planetary events contain
two types of information that preferentially enable them to meet this
criterion, which point-lens events generally lack, and if we did not
remove this information prior to testing against
Equation~(\ref{eqn:D8.3}), we would bias the sample toward planets.
First, planetary events can contain additional ground-based,
annual-parallax information due to sharp features in the light curve.
This was first clearly noted by \citet{mb09266} for MOA-2009-BLG-266.
Second, planetary events lead to measurement of (or strong constraints
upon) $\theta_\e$, which obviously aids in the determination of
$\pi_\rel=\theta_\e\pi_\e$.

Hence, we must determine not $\sigma(D_{8.3})$ for the actual event
but the same quantity for an analog event without planets. In Section
\ref{sec:fakedata}, we describe the creation of this analog
dataset. We present a qualitative analysis of the parallax signal in
Section \ref{sec:heuristic}, and the final determination in Section
\ref{sec:full}.

\subsubsection{The Modified Ground-based Dataset}
\label{sec:fakedata}

To construct this analog data set, we first take note of the 11 best
fit geometric parameters in the 2-orbital-parameter models
$a_k=(t_0,u_0,t_\e,\rho,\pi_{\e,N},\pi_{\e,E},s_0,q,\alpha_0,ds/dt,d\alpha/dt)$.
and the $(f_s,f_b)_i$ for each observatory $i$.  There are actually
four such solutions, a point to which we return immediately below.
Next we calculate the model magnification $A_{i,j}$ at each time
$t_{i,j}$ at observatory $i$, and so the model magnitudes $I_{i,j,\rm
  mod}=18 - 2.5\log(f_{s,i}A_{i,j}+f_{b,i})$ and hence the
corresponding residuals $r_{i,j} = I_{i,j,\rm obs}-I_{i,j,\rm
  mod}$. We then ignore the last five parameters, and calculate model
magnifications $A^\prime(t_{i,j};a_k, k=1, \ldots 6)$ and so model
magnitudes $I^\prime_{i,j,\rm mod}=18 -
2.5\log(f_{s,i}A^\prime_{i,j}+f_{b,i})$ and so simulated point-lens
``observations'' $I^\prime_{i,j,\rm obs} = I^\prime_{i,j,\rm mod} +
r_{i,j}$.  Because there are four parallax solutions, there are
actually four such models $I^\prime_{\rm mod}$.  This could create a
problem, in principle, because to mimic the treatment of point-lens
events, we must inject a single fake light curve into the
\citet{zhu17} procedures, not four.  We can exclude two of these
solutions, as discussed in Section~\ref{sec:twosol}.  (In any case,
the models derived from the $(+,+)$ and $(+,-)$ solutions are nearly
identical, as are those derived from the $(-,+)$ and $(-,-)$
solutions.)  To construct the fake data set, we average the two
models, $(-,+)$ and $(+,-)$.  We find that these models differ from
the average by a maximum of 0.0025 mag, and that, with the exception
of a 3-day interval near peak, the difference is less than 0.0015 mag.
These differences are far below the level at which the parallax
information contained in the ground-based light curve can be
significantly corrupted by averaging.

In principle, we should apply this procedure to {\it Spitzer} data as
well. However, since the restricted (``objective'') subset of the {\it
  Spitzer} data that are modeled in this test are not affected by the
presence of the planet, we simply use the {\it Spitzer} data as
observed.

\subsubsection{Heuristic Analysis of Restricted ``Objectively Selected'' {\it Spitzer} Data}
\label{sec:heuristic}

The objectively selected {\it Spitzer} data constitute no more than a
fragment of a light curve: 9 points spanning only 9 days of the
decline of a $t_\e\sim 100\,$day event. Yet, as we show below, because
the event is highly magnified, these few points are sufficient to
satisfy Equation~(\ref{eqn:D8.3}). \citet{gould12} showed that if a
high-magnification event is observed from space at the moment of its
ground-based peak, then the amplitude of $\bpi_\e$ can be determined
from that single point (together with a point at baseline to constrain
$f_b$), although the direction of $\bpi_\e$ will then be completely
undetermined.  In the restricted {\it Spitzer} dataset, the first
point is only 3.9 days after peak. The ground-based magnification at
this first {\it Spitzer} point is still quite high $A\sim 25$, while
the baseline flux is relatively well constrained by the 9 days of data
that are available. Thus, we might expect a reasonable measurement of
$\pi_\e$, especially since we also have color-color constraints.

As was the case for enforcing the \citet{yee15} protocols
(Section~\ref{sec:yee_proc}), we must be meticulous about mimicking
what would have happened if there had been no subjective selection.
There are several differences between the analysis of the full {\it
  Spitzer} light curve and the ``objective'' {\it Spitzer} light
curve.  First, by chance, when the errors are renormalized to enforce
$\chi^2/{\rm dof}=1$ on the restricted data set, the renormalization
factor is greater by a factor 1.13.  Then, in contrast to the
well-localized solution for the full data set, this fragment does not
yield a well-localized $(t_0,u_0)_{\rm sat}$, even when the $t_\e$ and
$f_s$ constraints are imposed. We find that at $1\,\sigma$,
$7576<t_0<7581$ and $0.006<u_0<0.083$.  Using $D_\perp=1.3\,\au$, this
yields ranges of $\Delta\pi_{\e,E}= 0.041$ and $\Delta\pi_{\e,N}=
0.059$, respectively, which would seem to imply a very poorly measured
$\pi_\e$.  Nevertheless, the values of $t_0$ and $u_0$ are highly
anti-correlated, so that the full $1\,\sigma$ range of $\pi_\e$ is
only 0.050--0.077.  This correlation, and corresponding good
constraint on the magnitude of $\bpi_\e$ (even though the direction is
basically unconstrained), is exactly what one would expect based on
the modified \citet{gould12} argument given above.

\subsubsection{Full Analysis of Restricted ``Objectively Selected'' {\it Spitzer} Data}
\label{sec:full}

Quantitatively, we generate MCMC samples according to the prescription
outlined in
Section~\ref{sec:heuristic} and analyze these using exactly the same
software as was employed by \citet{zhu17} for the {\it Spitzer}
point-lens sample that was used to establish the criterion
$\sigma(D_{8.3})< 1.4\,\kpc$.  We find
\begin{equation}
D_{8.3} = 6.9\pm 0.8\,\kpc
\qquad ({\rm Simulated\ point\ lens;\ restricted}\ Spitzer\ {\rm data}).
\label{eqn:sigma8.3}
\end{equation}
Hence, even with the restricted {\it Spitzer} data set, OGLE-2016-BLG-1190
easily satisfies the \citet{zhu17} criterion.

\section{Discussion}
\label{sec:discuss}

\subsection{First {\it Spitzer} Microlens Planet in the Bulge?}
\label{sec:bulge_planet}

Figure~\ref{fig:cumul} compares the $D_{8.3}$ cumulative distribution
of the planet sensitivities from \citet{zhu17} with that of the four
published {\it Spitzer} planets, all of which satisfy the
\citet{yee15} and \citet{zhu17} protocols.  This comparison cannot be
used to derive rigorous conclusions because the \citet{zhu17} sample
of high-cadence events from the 2015 season is not necessarily
representative of the full sample of events in which the four planets
were detected.  Nevertheless, Figure~\ref{fig:cumul} suggests that
there is not yet any clear difference in $D_{8.3}$ between the
planetary sample and the underlying population.  In particular, this
Figure does not depend in any way on our tentative conclusion that the
lens is in the bulge.

We note, however, that the experiment defined by \citet{yee15} can be
used to more finely distinguish between disk and bulge planet
populations than a simple Kolmogorov-Smirnov test comparing
distributions of planets detected with underlying planet
sensitivities.  The planet sample differs from the planet-sensitivity
sample in that for the former, $\theta_\e$ is usually measured or
constrained, providing additional information not available to the
underlying sample of events.  Moreover, since the planet sample is
substantially smaller, it can be subject to (typically
expensive) additional observations that can decisively resolve
ambiguities in disk/bulge classification.  Indeed, of the four
published {\it Spitzer} planets, only OGLE-2016-BLG-1190 is in need of
such further classification.

On the other hand, because the planet-sensitivity sample is much
larger, there is no need to determine which lenses individually lie in
the bulge to know the fraction of all the sensitivity that lies in the
bulge.  As demonstrated by \citet{zhu17}, the sensitivity lying at
intermediate $D_{8.3}$ can be divided between disk and bulge based on
Galactic models.

\subsection{Is OGLE-2016-BLG-1190Lb Really A Planet?}
\label{sec:really_planet}

If one wishes to answer this question according to the conventional
definition of ``not massive enough to burn deuterium'', then the
answer is ``perhaps, and a decisive answer may be obtained once
the host is resolved and its mass is estimated more precisely.''

However, as we have argued in Section~\ref{sec:intro}, this is
not a particularly interesting scientific question.  Rather,
we would like to know whether this ``planet'' (or ``brown dwarf'')
formed within the disk of its host (like ``planets'') or by
fragmentation (like ``stars'').  It is only the extreme paucity
of means to address this question that prevents it from being
asked in this form.  Nevertheless, this manner of posing the
question does react back upon the basic program for measuring
the ``Galactic distribution of planets''.

Figure~\ref{fig:massmass} shows the host and planet masses of the four
published {\it Spitzer} planets.  One sees immediately that the host
masses vary by a factor 10 and the planet masses vary by a factor of
3000.  This is far from a homogeneous sample, which certainly involves
very different formation processes.  The implicit assumption of the
``Galactic distribution'' program is that however heterogeneous these
processes are, one can at least tell whether the ensemble of formation
mechanisms is more efficient in the disk or bulge.

However, if the sample is being contaminated at the high-mass end by
``failed stars'', and if this contamination is different between the
disk and bulge, then it could make the comparison much more difficult.
In particular, \citet{thompson13} has argued that gas-giant planets
could be suppressed in the bulge due to the high-radiation environment
at the time of formation.  Probably such a mechanism would not
similarly suppress fragmentation leading to extreme super-Jupiter
``planets'' that had masses similar to OGLE-2016-BLG-1190Lb.  An
important signature of such suppression would then be a paucity of
Saturn, Jupiter, and few-Jupiter-mass planets in the bulge.  If this
paucity were complemented by a significant population of extreme
super-Jupiters in the bulge, it could indicate that the latter were
generated by a different formation mechanism.

\subsection{Toward Full Kepler Orbits: A Key Test of Microlens Orbital Motion}
\label{sec:fullkepler}

While it did not prove possible to completely measure the full set of
Kepler parameters (which would have provided the first microlensing
measurement of a planet's eccentricity), the attempt to do so
unexpectedly led to the first test of microlensing orbital motion
measurements.  Recall from Section~\ref{sec:intro} that the first
microlensing orbital motion measurement (for MACHO-97-BLG-41) appeared
to be possible only because the source happened to pass over an
outlying caustic at a different time and at a different angle than was
``predicted'' based on the model of the light curve in the region of
the central caustic, which occurred five weeks later.  Hence, such
orbital-motion measurements were regarded at the time as requiring
very exceptional circumstances.  Subsequently, many orbital motion
measurements were reported based solely on the source passage over the
central caustic.  For example, \citet{ob05071b} reported such a
measurement for OGLE-2005-BLG-071, based on a light curve with two cusp
approaches separated by just three days.  However, in contrast to
MACHO-97-BLG-41, there is no intuitive way to see from the light curve
that orbital motion is really being detected, and there has never been
a clear test that these ``non-intuitive'' orbital-motion parameters
are being correctly measured.  In lieu of rigorous tests, one can
check whether the ratio of transverse kinetic to potential energy
$\beta\equiv {\rm (KE/PE)}_\perp$ \citep{ob05071b} satisfies the
physical requirement $\beta<1$, and more generally whether $\beta$
lies in a plausible range.  However, such tests are only qualitative,
and in particular, if $\beta\ll 1$ this does not prove that there is
any problem in the measurement, only that the system is viewed from a
relatively unlikely perspective\footnote{For both of the surviving
``large parallax'' solutions of OGLE-2016-BLG-1190, $\beta\simeq 0.25$.}.

\begin{figure}
\plotone{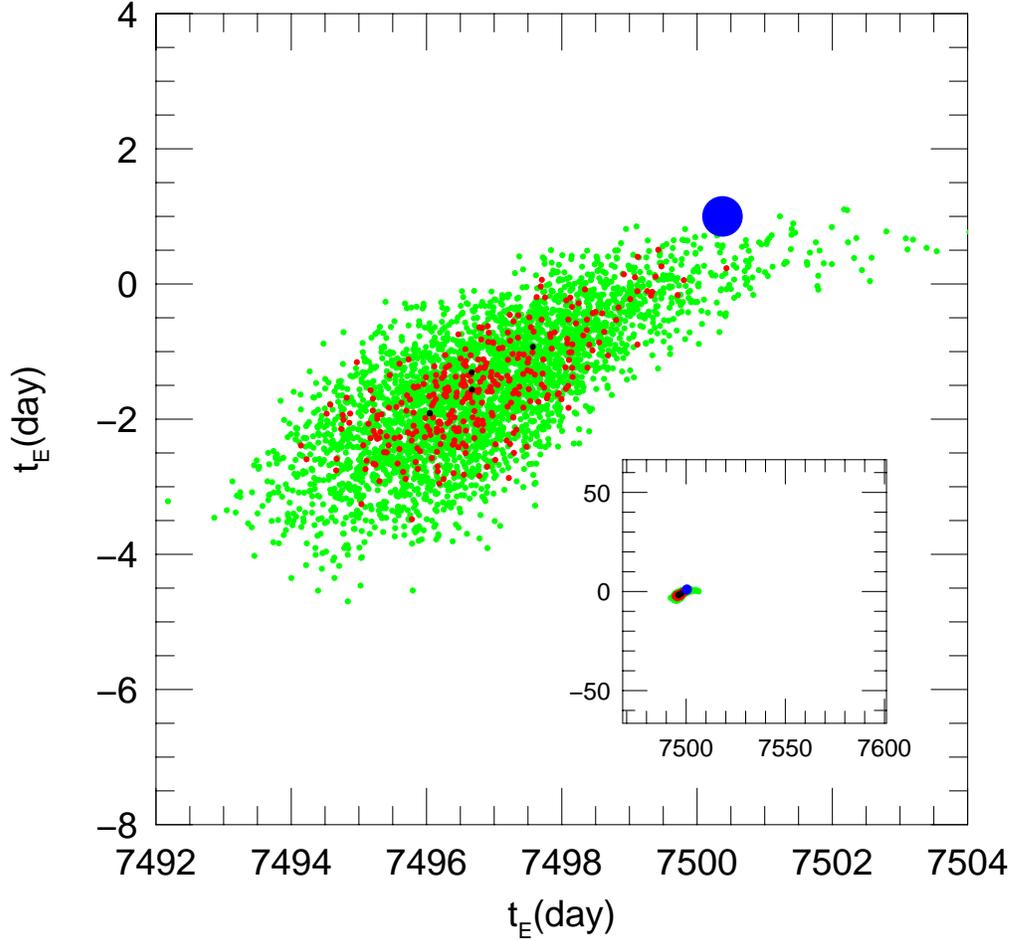} \caption{``Predicted'' (small circles) versus
``observed'' (large blue circle) position of source crossing of
planetary caustic within the Einstein ring.  The predictions are
from an MCMC chain created by fitting both ground-based and {\it
Spitzer} data to a model with linearized orbital motion, but with
the data points in the neighborhood of the observed planetary
caustic crossing omitted.  Points are colored by (black, red, green)
for $\Delta\chi^2 <(1,4,9)$. The abscissa of the prediction for each
chain element is the time that the source should have crossed the
center of the planetary caustic.  The ordinate is that of the center
of the caustic at this time, multiplied by $t_\e$.  The abscissa of
the ``observed'' position is the mid-time of the two caustic
crossings shown in the lower-left panel of Figure~\ref{fig:lc}.  The
ordinate is that of the source position at this time, multiplied by
$t_\e$. Even without any ``knowledge'' of the source crossing, the
model predicts its position very accurately.  Inset shows zoom-out
on the same scale as Figure~\ref{fig:geolin}. } \label{fig:pred}
\end{figure}

\begin{figure}
\plotone{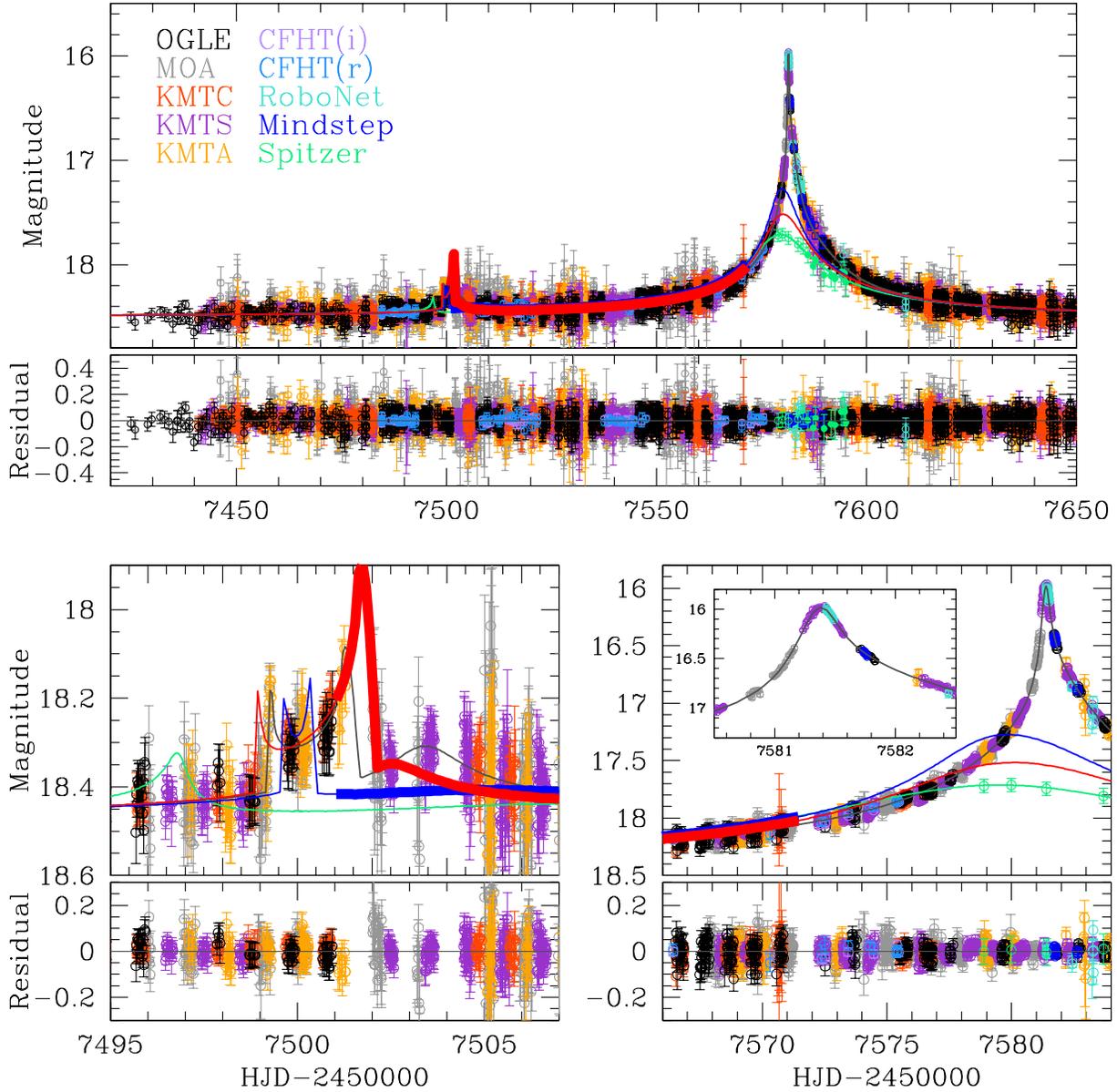} \caption{Similar to Figure~\ref{fig:lc} except
that the predictions for {\it Kepler} ``K2'' observations
\citep{henderson16} are shown for the $(+,+)$ (red) and $(+,-)$
(blue) solutions.  The time intervals when data were actually taken
are shown in thick lines, while the times with no data are shown in
thin lines.  The large-parallax solutions [$(-,+)$ and $(+,-)$]
predict a fainter peak from K2. See Equation~(\ref{eqn:arat}).
Unfortunately, K2 observations ended nine days before peak (lower
right panel).  Nevertheless, the two models predict radically
different light curves for the planetary caustic crossing 80 days
earlier.  See text. } \label{fig:k2_lc}
\end{figure}

\begin{figure}
\plotone{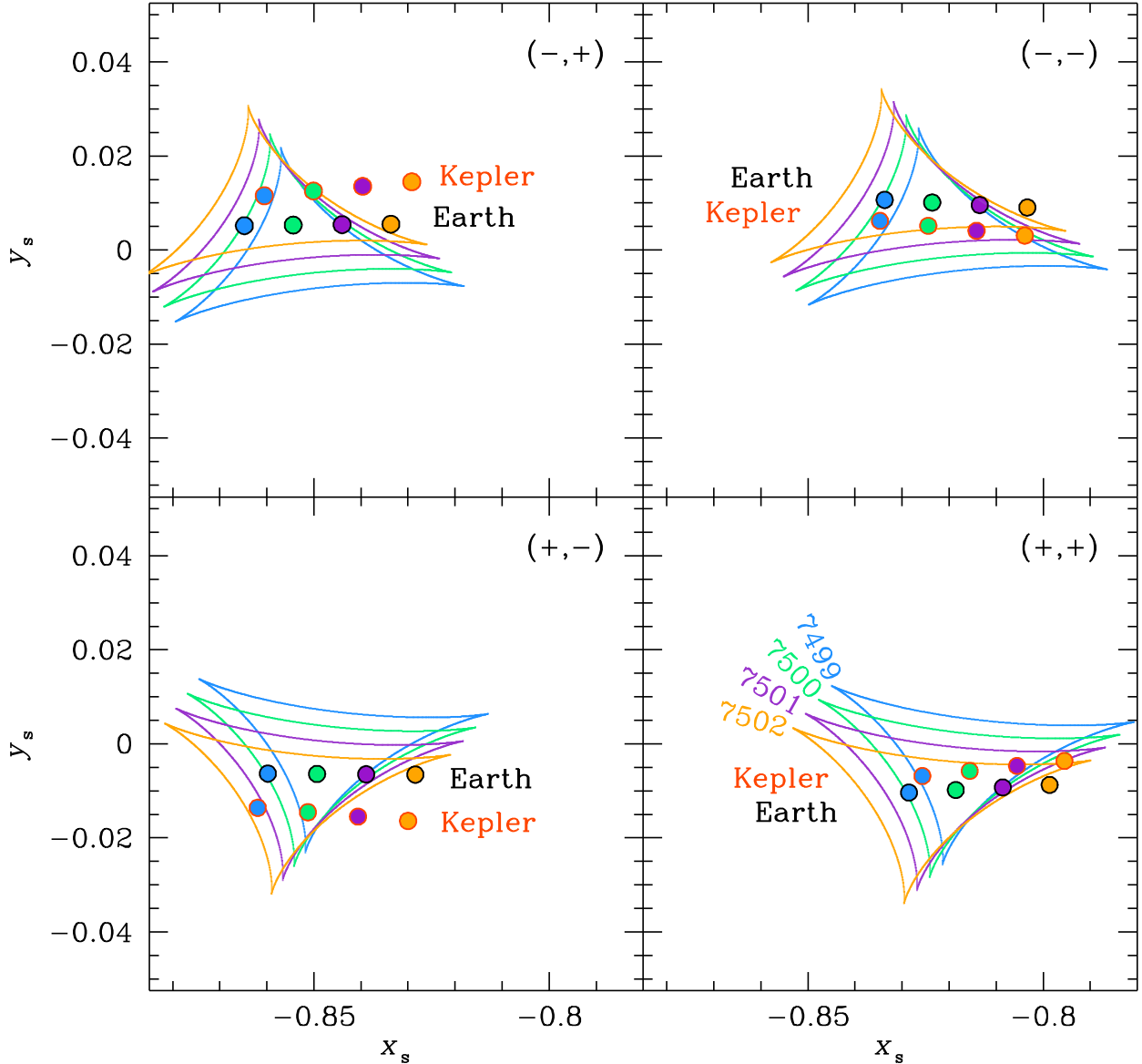}
\caption{Planetary-caustic geometries for {\it Kepler} ``K2'' and
Earth-based observations for the four solutions that are degenerate
based on Earth-based and {\it Spitzer} data alone.  For each day
HJD$^\prime=(7499, \ldots ,7502)$, the caustic and the projected
positions of the source as seen from {\it Kepler} and Earth are shown
in the same color.  In each of the four panels, the Earth-viewed source
at a particular time is in a very similar position relative to the caustic,
even though the caustic itself does not have the same position or orientation
in the Einstein ring.  This simply reflects that the model must match
the ground-based data seen in Figures~\ref{fig:lc},
\ref{fig:lc_zoom},
and \ref{fig:k2_lc}.
The vector offset between the Earth and {\it Kepler}
positions is nearly constant within a given panel because their projected
separation ${\bf D}_\perp$ barely changes during this interval.  Because
$\bpi_\e$ is quite precisely determined within each of the four
solutions, the {\it Kepler} trajectory through the caustic is likewise
well determined.  But because of the opposite sign $\pi_{\e,N}$ in,
e.g., the $(-,+)$ and $(-,-)$ solutions, the ${\it Kepler}$-viewed source
passes closer to the tip of the caustic in the first case (and so leaves
the caustic before the first K2 data point at $\sim 7501$) whereas
it passes closer to the base of the caustic in the second case (and
so exits the caustic after the start of K2 observations).  Hence
the prediction of a dramatic difference for the ``large parallax'' (left
panels) and ``small parallax'' (right panels) solutions is robust.
}
\label{fig:k2_caust}
\end{figure}

\begin{figure}
\plotone{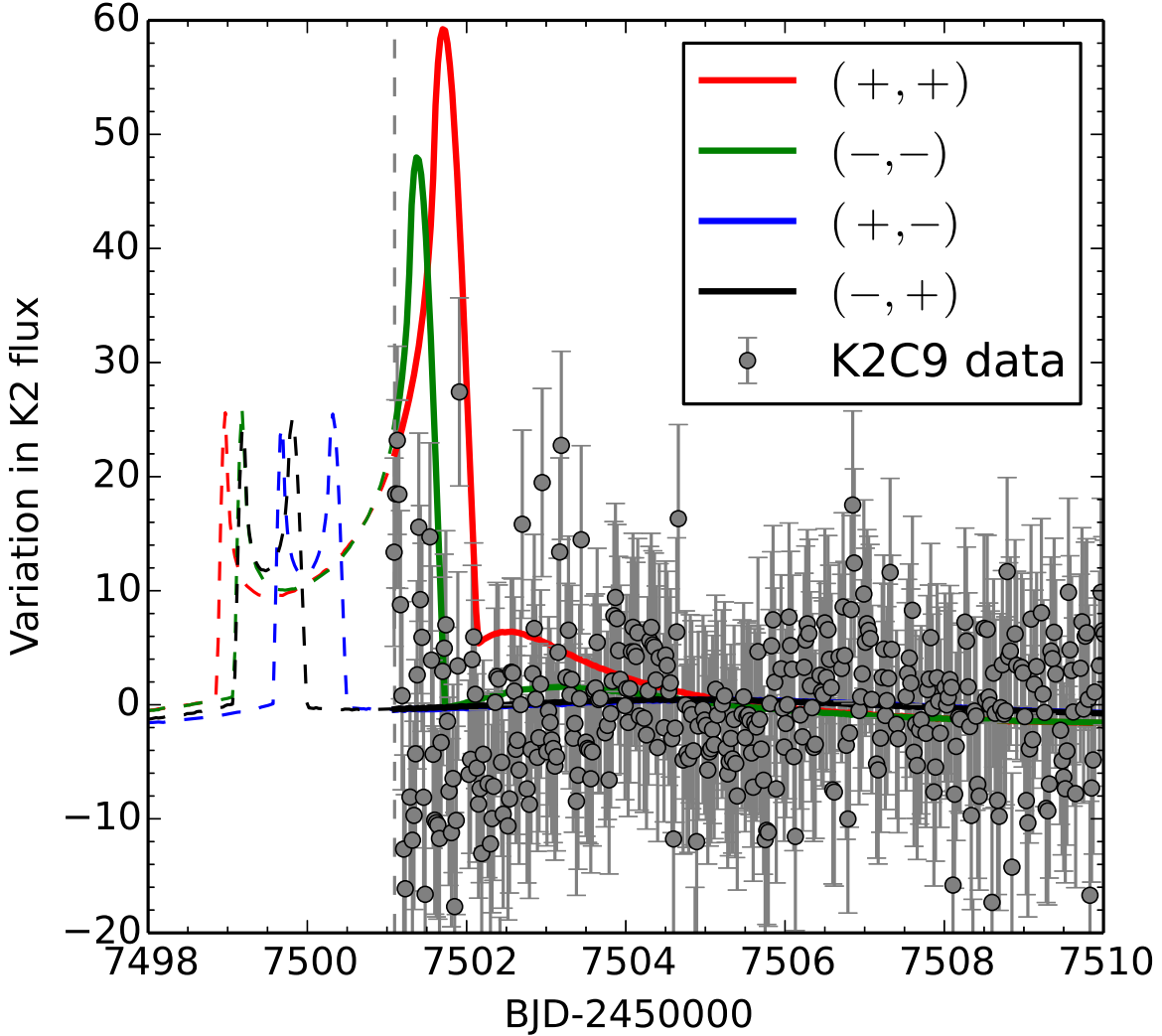}
\caption{Observed (black points) K2 light curve compared to four microlens
parallax model that would be degenerate based on ground-based and
{\it Spitzer} data alone.  The data do not show any indication of a spike
at the predicted caustic exits of the two ``small parallax'' solutions
[$(-,-)$ and $(+,+)$].  The times of these exits are very well predicted,
as discussed in Section~\ref{sec:k2} and Figure~\ref{fig:k2_caust}.
In order to determine the amplitude of these spikes, one must derive
the source flux in the {\it Kepler} band $K_p$.  This is done essentially
by a $VIK_p$ color-color relation, with a small correction term based
on the extinction, yielding $f_{s,Kepler}= 6.1$.  Hence, e.g., the
red $(+,+)$ curve drops by $\Delta A=9.8$ magnification units between
the peak and the post-caustic ``baseline''.
The uncertainty in this transformation is 0.1 mag,
implying a $\sim 10\%$ uncertainty in the height of the spikes, which
does not impact the robustness of the rejection of  the ``small parallax''
models. The $(+,+)$ and $(+,-)$ solutions are shown on a magnitude
scale in Figure~\ref{fig:k2_lc}, with the regions probed by K2 data
(see this figure) highlighted in boldface.
}
\label{fig:k2obs}
\end{figure}

\begin{figure}
\plotone{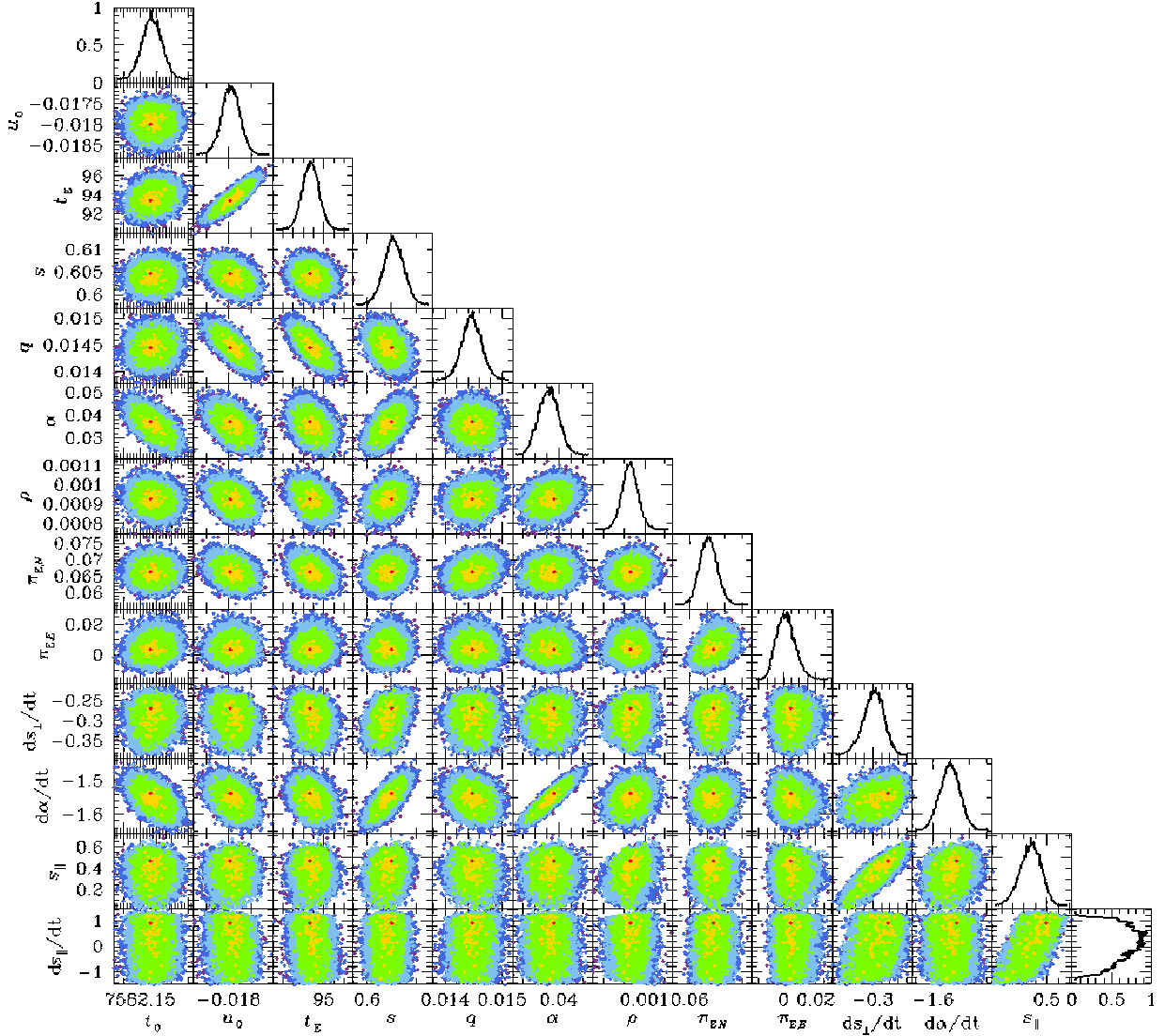}
\caption{Scatter plots of pairs of all 13 microlensing parameters from the
complete orbital solution for the $(-,+)$ minimum.  The plots
for the $(+,-)$ solution (which has a worse fit by
$\Delta\chi^2=4$) look qualitatively similar.  (Red, yellow, green, cyan)
 points are within $\Delta\chi^2<(1,4,9,16)$ of the minimum.  All parameters
are relatively well-constrained except $ds_\parallel/dt$, i.e., the line of
sight velocity of the planet in units of Einstein radii per year.
}
\label{fig:mptri}
\end{figure}

\begin{figure}
\plotone{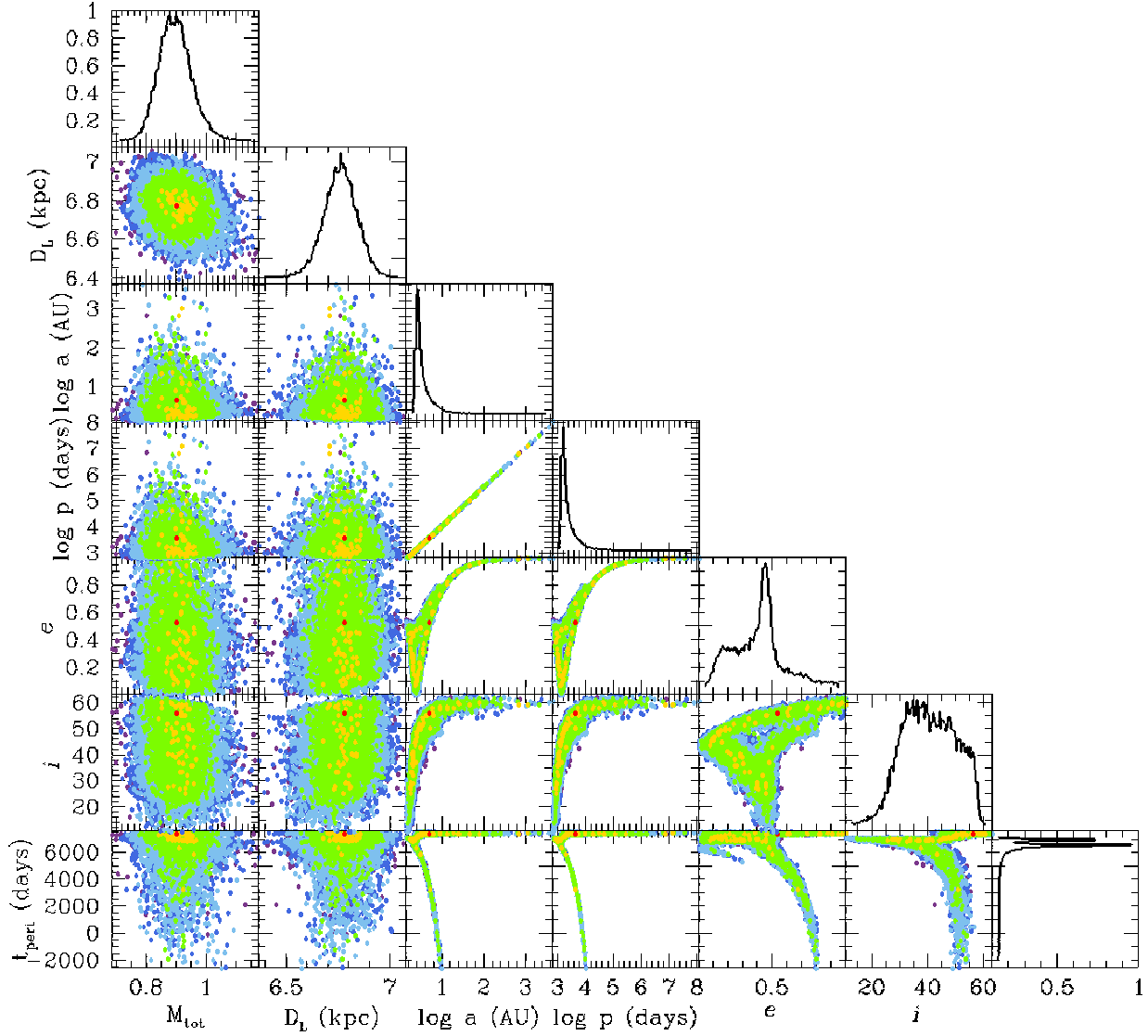}
\caption{Scatter plots of pairs of 6 Kepler parameters that are derived
from the chain of the
complete orbital solution for the $(-,+)$ minimum.
(Red, yellow, green, cyan)
 points are within $\Delta\chi^2<(1,4,9,16)$ of the minimum.
The Kepler parameters period $P$, eccentricity $e$, inclination $i$,
and time of periastron $t_{\rm peri}$ are confined to an essentially
one-dimensional sub-space.
}
\label{fig:mpphys}
\end{figure}

\begin{figure}
\plotone{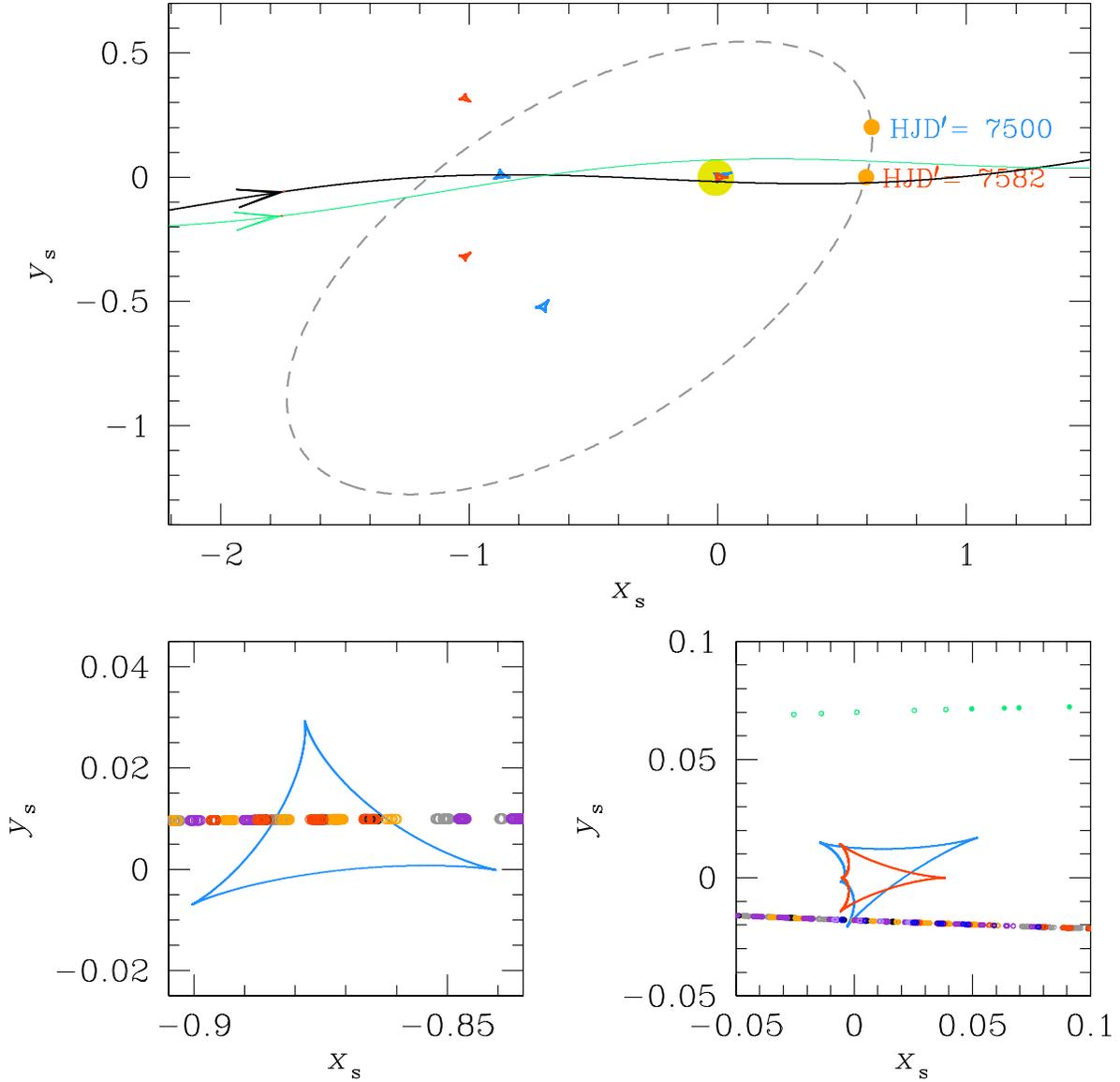} \caption{Geometries of the source and lens system
for the $(-,+)$ solution.
The dashed gray line shows the planet's orbit for the best-fit
model with its position at the times of the two caustic crossings shown
by orange dots.  However, the parameters such
as the orbital inclination and eccentricity have large
uncertainties, and thus there are many possible orbital geometries.
The caustic structure at the first of these
epochs is shown in blue and at the second in red.  The trajectories
of the source position through the Einstein ring as seen from {\it
Spitzer} and Earth are shown in green and black, respectively. Their
``waviness'' reflects the heliocentric orbital motion of these two
observatories. Epochs of observations are shown by small circles,
using same color scheme as in Figure~\ref{fig:lc}. The yellow circle
at the position $(X_s,Y_s)=(0,0)$ shows the position of lens star.
The corresponding $(+,-)$ diagram looks essentially identical, but
inverted with respect to the $x$-axis.
}
\label{fig:geo}
\end{figure}

\begin{figure}
\plotone{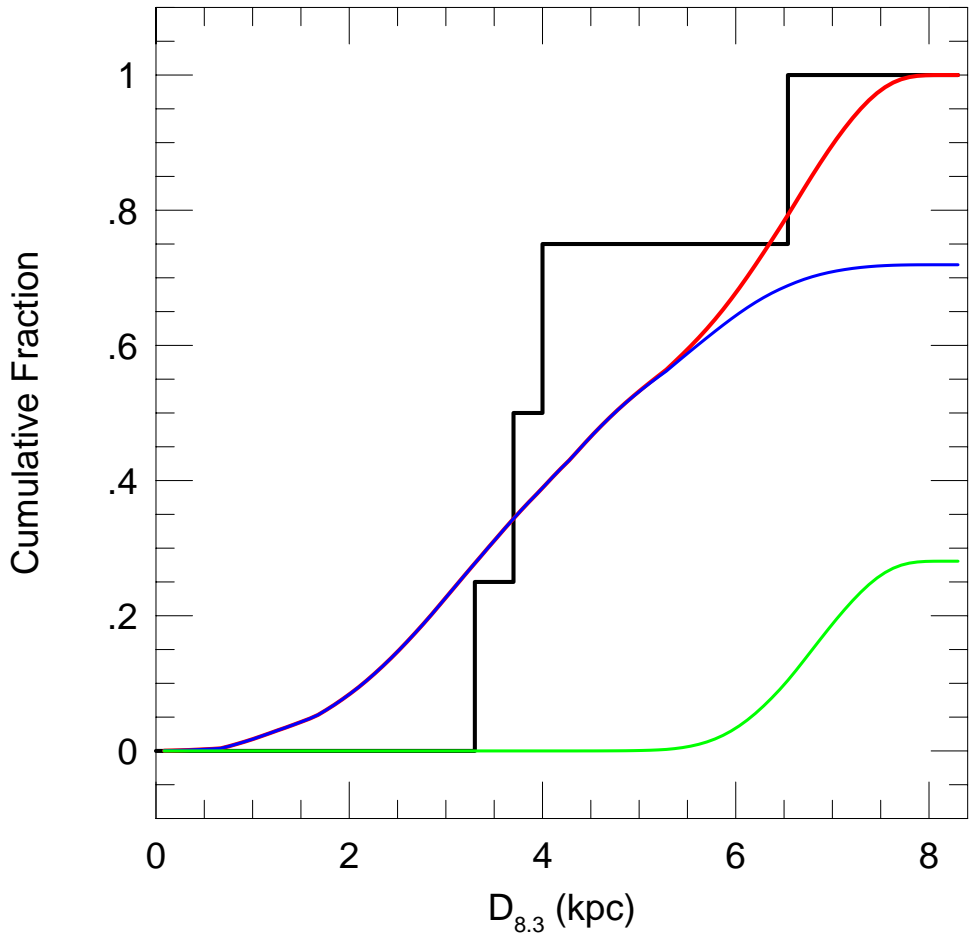} \caption{ Cumulative distribution in ``distance
indicator'' $D_{8.3}$ of planet sensitivities derived from 41
high-cadence 2015 {\it Spitzer} events \citep{zhu17} compared to
that of the four {\it Spitzer} planets published to date (black).
The comparison is only meant to be illustrative because the
\citet{zhu17} sample is not necessarily representative of the full
{\it Spitzer} sample.  Nevertheless, the addition of
OGLE-2016-BLG-1190Lb (this paper) at $D_{8.3}=6.5\,\kpc$ breaks the
previous pattern of relatively nearby lenses established by
OGLE-2014-BLG-0124 \citep{ob140124,ob140124b}, OGLE-2015-BLG-0966
\citep{ob150966}, and OGLE-2016-BLG-1195 \citep{ob161195}. The full
cumulative distribution for the high-cadence sample (red), is
divided into disk (blue) and bulge (green) lenses. }
\label{fig:cumul}
\end{figure}

\begin{figure}
\plotone{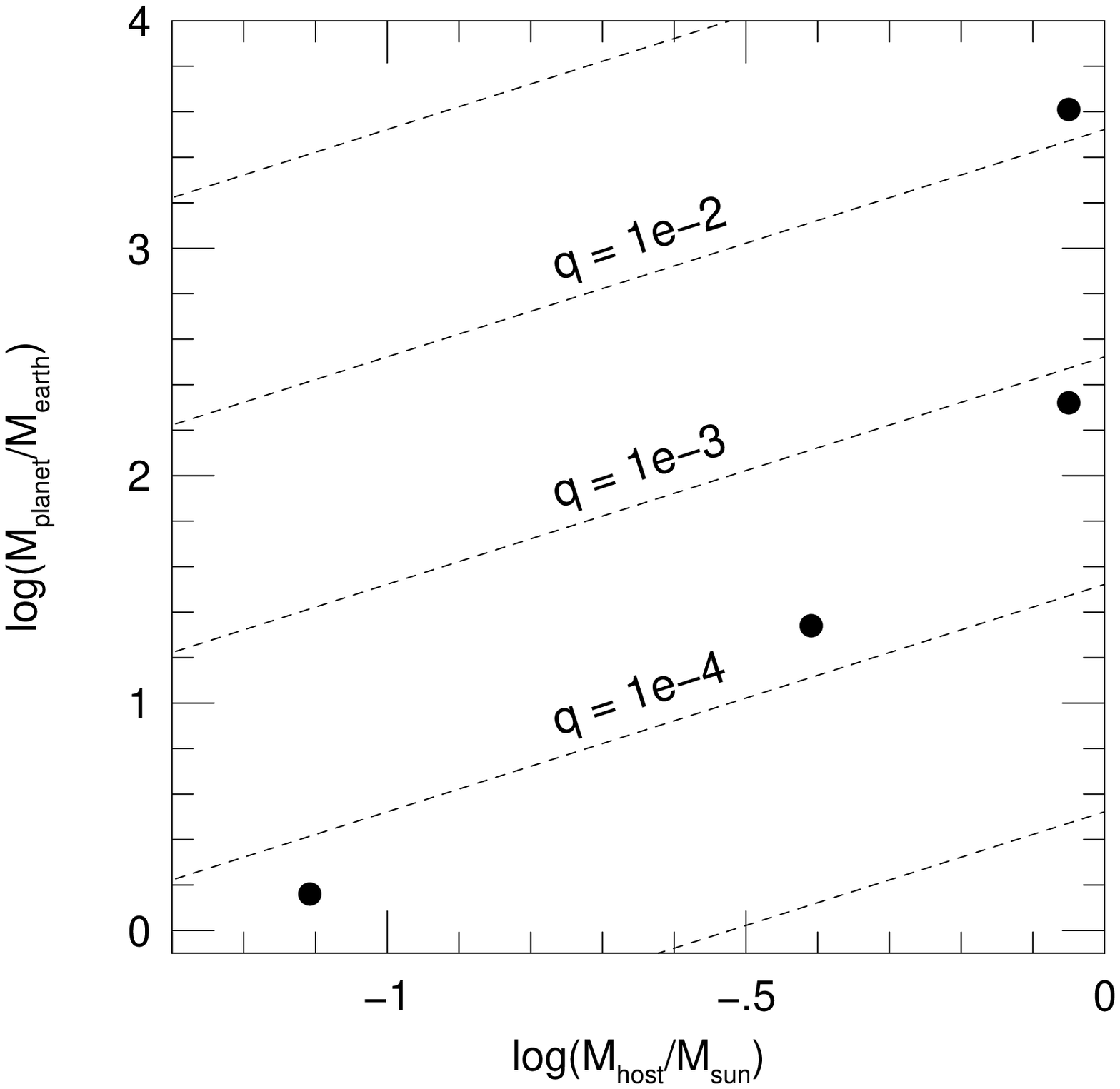} \caption{log($M_{\rm planet}$) vs log($M_{\rm
host}$) for the four published {\it Spitzer} microlensing planets,
OGLE-2014-BLG-0124 \citep{ob140124,ob140124b}, OGLE-2015-BLG-0966
\citep{ob150966}, OGLE-2016-BLG-1195 \citep{ob161195}, and
OGLE-2016-BLG-1190 (this work).  The dashed diagonal lines indicate
constant mass ratio $q=M_{\rm planet}/M_{\rm host}$.  This very
small sample already contains an extreme diversity of objects,
spanning factors of 10 in $M_{\rm host}$, 300 in $q$ and 3000 in
$M_{\rm planet}$. } \label{fig:massmass}
\end{figure}

In the case of OGLE-2016-BLG-1190, however, we do have such a rigorous
test.  We showed in Figure~\ref{fig:pred} that
models derived after excluding the data within $\pm 10\,$days
of the planetary caustic yielded a precise measurement of the orbital
motion that is ``confirmed'' by the actual orbital motion from the
full data set.  Moreover, we showed in Table~\ref{tab:deletedata}
that models derived from a wide variety of subsets of the actual
data yield orbital motion estimates that are consistent with the
true ones, within their own errors.  Although
events that exhibit both planetary (or more generally outlying) and
central caustic crossings are rare, we suggest that these may provide
an excellent set of tests for the accuracy of orbital-motion
measurements from central caustic crossings.

\acknowledgments
Work by WZ, YKJ, and AG were supported by AST-1516842 from the US NSF.
WZ, IGS, and AG were supported by JPL grant 1500811.
This research has made use of the KMTNet system operated by the Korea
Astronomy and Space Science Institute (KASI) and the data were obtained at
three host sites of CTIO in Chile, SAAO in South Africa, and SSO in
Australia.
Work by C.H. was supported by the grant (2017R1A4A101517) of
National Research Foundation of Korea.
The OGLE project has received funding from the National Science Centre,
Poland, grant MAESTRO 2014/14/A/ST9/00121 to AU.
The MOA project is supported by JSPS KAKENHI Grant Number JSPS24253004, JSPS26247023, JSPS23340064, JSPS15H00781, and JP16H06287.
Work by CR was supported by an appointment to the NASA Postdoctoral Program
at the Goddard Space Flight Center, administered by USRA through a contract
with NASA.
Work by YS was supported by an appointment to the NASA Postdoctoral Program at the Jet Propulsion Laboratory,
California Institute of Technology, administered by Universities Space Research Association
through a contract with NASA.
The United Kingdom Infrared Telescope (UKIRT) is supported by NASA and
operated under an agreement among the University of Hawaii, the University
of Arizona, and Lockheed Martin Advanced Technology Center; operations are
enabled through the cooperation of the Joint Astronomy Centre of the Science
and Technology Facilities Council of the U.K.
We acknowledge the support from NASA HQ for the UKIRT observations in connection with $K2$C9.
This research uses data obtained through the Telescope Access Program
(TAP), which has been funded by the National Astronomical
Observatories of China, the Chinese Academy of Sciences (the Strategic
Priority Research Program ``The Emergence of Cosmological Structures''
Grant No. XDB09000000), and the Special Fund for Astronomy from the
Ministry of Finance. Based on observations obtained with
MegaPrime/MegaCam, a joint project of CFHT and CEA/DAPNIA, at the
Canada-France-Hawaii Telescope (CFHT) which is operated by the
National Research Council (NRC) of Canada, the Institut National des
Science de l'Univers of the Centre National de la Recherche
Scientifique (CNRS) of France, and the University of Hawaii.  This
work was performed in part under contract with the California
Institute of Technology (Caltech)/Jet Propulsion Laboratory (JPL)
funded by NASA through the Sagan Fellowship Program executed by the
NASA Exoplanet Science Institute. Work by MTP and BSG was supported by
NASA grant NNX16AC62G.  This work was partly supported by the National
Science Foundation of China (Grant No. 11333003, 11390372 to SM).
KH acknowledges support from STFC grant ST/M001296/1.
This work makes use of observations from the LCOGT network, which
includes three SUPAscopes owned by the University of St. Andrews. The
RoboNet programme is an LCO Key Project using time allocations from
the University of St Andrews, LCOGT and the University of Heidelberg
together with time on the Liverpool Telescope through the Science and
Technology Facilities Council (STFC), UK. This research has made use
of the LCO Archive, which is operated by the California Institute of
Technology, under contract with the Las Cumbres Observatory.
Work by S.R. and S.S. is supported by INSF-95843339

\begin{deluxetable}{lccc}
\tablecolumns{4} \tablewidth{0pc}\tablecaption{\textsc{Observatory}}
\tablehead{ \colhead{Data set} & \colhead{Number} &
\colhead{$\chi^2$} & \colhead{Filter}} \startdata
  OGLE              & 3293 & 3290.161 & I \\
  KMTC (BLG03)      & 1510 & 1508.821 & I \\
  KMTC (BLG43)      & 1437 & 1435.652 & I \\
  KMTS (BLG03)      & 1770 & 1768.444 & I \\
  KMTS (BLG43)      & 1713 & 1712.087 & I \\
  KMTA (BLG03)      & 1108 & 1107.140 & I \\
  KMTA (BLG43)      & 1136 & 1135.246 & I \\
  MOA               & 2089 & 2088.061 & RI \\
  MiNDSTEp          & 37   & 36.908   & I\\
  RoboNet           & 40   & 40.068   & i \\
  CFHT              & 67   & 66.870   & i \\
  CFHT              & 74   & 73.962   & r \\
  {\it Spitzer}     & 14   & 10.453   & L \\
\enddata
\label{tab:observe}
\end{deluxetable}

\clearpage

\begin{deluxetable}{lcccc}
\tablecolumns{6} \tablewidth{0pc}\tablecaption{\textsc{Best-fit
solutions for parallax+orbital motion (2-parameters) models}}
\tablehead{ \colhead{Parameters } & \colhead{$(-,+)$}
&\colhead{$(+,-)$}& \colhead{$(+,+)$}&\colhead{$(-,-)$}} \startdata
  $\chi^2/\rm{dof}$               &14283.479/14252       &14292.586/14252       &14296.670/14252        &14302.523/14252       \\
  $t_0$ $(\rm{HJD}^{\prime})$     &7582.161 $\pm$ 0.007  &7582.160 $\pm$ 0.007  &7582.167 $\pm$ 0.007   &7582.167 $\pm$ 0.007  \\
  $u_0$ $(10^{-2})$               &-1.747 $\pm$ 0.023    &1.717 $\pm$ 0.023     &1.667 $\pm$ 0.021      &-1.667 $\pm$ 0.022   \\
  $t_{\rm E}$ $(\rm{days})$       &95.747 $\pm$ 0.958    &97.354 $\pm$ 1.006    &100.161 $\pm$ 0.952    &100.192 $\pm$ 0.983    \\
  $s$                             &0.604 $\pm$ 0.002     &0.603 $\pm$ 0.002     &0.603 $\pm$ 0.002      &0.604 $\pm$ 0.002     \\
  $q$ $(10^{-2})$                 &1.414 $\pm$ 0.019     &1.393 $\pm$ 0.019     &1.360 $\pm$ 0.017      &1.354 $\pm$ 0.018     \\
  $\alpha$ $(\rm{rad})$           &0.033 $\pm$ 0.005     &-0.033 $\pm$ 0.005    &-0.028 $\pm$ 0.005     &0.030 $\pm$ 0.005     \\
  $\rho$ $(10^{-3})$              &0.908 $\pm$ 0.050     &0.873 $\pm$ 0.050     &0.860 $\pm$ 0.045      &0.868 $\pm$ 0.046     \\
  $\pi_{\rm{E},\it{N}}$           &0.065 $\pm$ 0.003     &-0.063 $\pm$ 0.002    &0.038 $\pm$ 0.002      &-0.037 $\pm$ 0.002    \\
  $\pi_{\rm{E},\it{E}}$           &0.004 $\pm$ 0.006     &0.011 $\pm$ 0.007     &0.008 $\pm$ 0.006      &0.011 $\pm$ 0.007     \\
  $ds/dt$ $(\rm{yr}^{-1})$        &-0.278 $\pm$ 0.018    &-0.286 $\pm$ 0.019    &-0.332 $\pm$ 0.018     &-0.320 $\pm$ 0.018    \\
  $d\alpha/dt$ $(\rm{yr}^{-1})$   &-1.417 $\pm$ 0.030    &1.402  $\pm$ 0.030    &1.394 $\pm$ 0.030      &-1.385 $\pm$ 0.030    \\
\enddata
\label{tab:ulens_2}
\end{deluxetable}

\clearpage
\begin{deluxetable}{lcc}
\tablecolumns{3} \tablewidth{0pc}\tablecaption{\textsc{Orbital Motion with Deleted Data}}
\tablehead{\colhead{Deleted
Data}&\colhead{$ds/dt$}&\colhead{$d\alpha/dt$}} \startdata
  None         &-0.278 $\pm$ 0.018 &-1.417 $\pm$ 0.030     \\
  7490---7510  &-0.211 $\pm$ 0.056 &-1.548 $\pm$ 0.056 \\
  7495---7567  &-0.234 $\pm$ 0.102 &-1.156 $\pm$ 0.115 \\
  7490---7567  &-0.195 $\pm$ 0.210 &-1.192 $\pm$ 0.282 \\
  7240---7567  &-0.361 $\pm$ 0.224 &-1.720 $\pm$ 0.423 \\
\enddata
\label{tab:deletedata}
\end{deluxetable}

\begin{deluxetable}{lcccc}
\tablecolumns{3} \tablewidth{0pc}\tablecaption{\textsc{Physical
properties from parallax+orbital motion (2-parameters) models}}
\tablehead{\colhead{Quantity}&\colhead{$(-,+)$}&\colhead{$(+,-)$}&\colhead{$(+,+)$}&\colhead{$(-,-)$}}
\startdata
  $M_1$ $[M_\sun]$   &0.91 $\pm$ 0.06   &0.92 $\pm$ 0.07  &1.60 $\pm$ 0.15  &1.60 $\pm$ 0.16       \\
  $M_2$ $[M_J]$      &13.51 $\pm$ 0.93  &13.52 $\pm$ 0.94 &22.78 $\pm$ 2.10 &22.74 $\pm$ 2.23    \\
  $D_{\rm L}$ [kpc]  &6.79 $\pm$ 0.10   &6.78 $\pm$ 0.10  &7.43 $\pm$ 0.10  &7.42 $\pm$ 0.11   \\
  $a_\bot$ [AU]      &2.03 $\pm$ 0.09   &2.04 $\pm$ 0.09  &2.28 $\pm$ 0.10  &2.28 $\pm$ 0.11    \\
  $\mu$ [mas/yr]     &1.88 $\pm$ 0.10   &1.88 $\pm$ 0.10  &1.86 $\pm$ 0.09  &1.86 $\pm$ 0.09    \\
\enddata
\label{tab:phys_2}
\end{deluxetable}

\begin{deluxetable}{lcc}
\tablecolumns{6} \tablewidth{0pc}\tablecaption{\textsc{Best-fit
solutions for parallax+orbital motion (4-parameters) models}}
\tablehead{ \colhead{Parameters} & \colhead{$(-,+)$}
&\colhead{$(+,-)$}} \startdata
  $\chi^2/\rm{dof}$               &14273.875/14250       &14277.653/14250         \\
  $t_0$ $(\rm{HJD}^{\prime})$     &7582.157 $\pm$ 0.007  &7582.154 $\pm$ 0.007    \\
  $u_0$ $(10^{-2})$               &-1.797 $\pm$ 0.023    &0.018 $\pm$ 0.0002      \\
  $t_{\rm E}$ $(\rm{days})$       &93.532 $\pm$ 0.891    &94.034 $\pm$ 0.934      \\
  $s$                             &0.604 $\pm$ 0.002     &0.604 $\pm$ 0.002       \\
  $q$ $(10^{-2})$                 &1.446 $\pm$ 0.019     &1.440 $\pm$ 0.019       \\
  $\alpha$ $(\rm{rad})$           &0.038 $\pm$ 0.005     &-0.039 $\pm$ 0.005      \\
  $\rho$ $(10^{-3})$              &0.930 $\pm$ 0.044     &0.908 $\pm$ 0.045       \\
  $\pi_{\rm{E},\it{N}}$           &0.067 $\pm$ 0.003     &-0.066 $\pm$ 0.002      \\
  $\pi_{\rm{E},\it{E}}$           &0.004 $\pm$ 0.006     &0.012 $\pm$ 0.006       \\
  $ds_{\bot}/dt$ $(\rm{yr}^{-1})$ &-0.265 $\pm$ 0.028    &-0.366 $\pm$ 0.027      \\
  $d\alpha/dt$ $(\rm{yr}^{-1})$   &-1.536 $\pm$ 0.030    &1.530  $\pm$ 0.030      \\
  $s_{\|}$                     &0.484 $\pm$ 0.101     &-0.011 $\pm$ 0.113      \\
  $ds_{\|}/dt$ $(\rm{yr}^{-1})$   &0.961 $\pm$ 0.662     &0.801  $\pm$ 0.715     \\
\enddata
\label{tab:ulens_4}
\end{deluxetable}



\begin{deluxetable}{lcccccc}
\tablecolumns{8} \tablewidth{0pc}\tablecaption{\textsc{$\chi^2$
comparison}} \tablehead{ \colhead{Model} & \colhead{dof}
&\colhead{$\chi^2_{\rm{gr}}$}&\colhead{$\chi^2_{\rm{sp}}$}&\colhead{$\chi^2_{\rm{tot}}$}
&\colhead{$\Delta\chi^2$}&\colhead{comp}}
\startdata
  Standard                       &14243 &17898.5 &0.   &17898.5 &---    &---      \\
  Orbital(2parameters)           &14241 &14283.7 &0.   &14283.7 &3614.8 &ground   \\
  Parallax+Orbital(2parameters)  &14252 &14273.1 &10.4 &14283.5 &10.6   &ground    \\
  Parallax+Orbital(4parameters)  &14250 &14263.4 &10.5 &14273.9 &9.6    &all      \\
\enddata
\label{tab:compchi2}
\end{deluxetable}


\begin{deluxetable}{lcc}
\tablecolumns{3} \tablewidth{0pc}\tablecaption{\textsc{Physical
properties from parallax+orbital motion (4-parameters) models}}
\tablehead{\colhead{Quantity}&\colhead{$(-,+)$}&\colhead{$(+,-)$}}
\startdata
  $M_1$ $[M_\sun]$              &$0.88_{-0.05}^{+0.06}$      &$0.89_{-0.06}^{+0.07}$     \\
  $M_2$ $[M_J]$                 &$13.38_{-0.82}^{+0.88}$     &$13.38_{-0.89}^{+0.97}$   \\
  $D_{\rm L}$ [kpc]             &$6.77_{-0.09}^{+0.08}$      &$6.74_{-0.09}^{+0.08}$     \\
  $a$ [AU]                      &$2.17_{-0.38}^{+1.87}$      &$2.04_{-0.43}^{+2.10}$     \\
  $P$ [yr]                      &$3.35_{-0.82}^{+5.19}$      &$3.05_{-0.89}^{+5.79}$     \\
  $\epsilon$                    &$0.42_{-0.23}^{+0.13}$      &$0.42_{-0.21}^{+0.11}$     \\
  $i$ [deg]                     &$41.20_{-10.29}^{+11.95}$   &$-39.93_{-16.45}^{+13.32}$     \\
  $t_{\rm peri}$ [HJD$^\prime$] &$6989.8_{-434.1}^{+423.2}$  &$7006.9_{-754.6}^{+486.5}$     \\
\enddata
\tablecomments{The inclination for the $(+,-)$ solution is shown 
as a negative number to make manifest the fundamental symmetry of the two
solutions. In standard notation, it would be $i+180^\circ\rightarrow
140.07^\circ$.} 
\label{tab:phys_4}
\end{deluxetable}

\begin{thebibliography}{99}


\bibitem[Albrow et al.(2000)]{mb9741}Albrow, M.\ Beaulieu, J.-P., Caldwell, J.A.R.,, et al.\ 2000, \apj, 534, 894

\bibitem[Albrow et al.(2009)]{albrow09}Albrow, M.\ D., Horne, K., Bramich, D.\ M., et al.\ 2009, \mnras, 397, 2099




\bibitem[Alard \& Lupton(1998)]{alard98} Alard, C. \& Lupton, R.H.,1998, \apj, 503, 325





\bibitem[Batista et al.(2014)]{mb11293b} Batista, V., Beaulieu, J.-P., Gould, A., et al. \apj, 780, 54


\bibitem[Beaulieu et al.(2017)] {ob140124b}Beaulieu, J.-P., Batista, V., Bennett, D.P., et al. 2017, \aj, in press, arXiv:1709.00806


\bibitem[Bennett et al.(1999)]{mb9741x}  Bennett, D.P., Rhie, D.P., Becker, A.C., et al.\ 1999, Natur, 402 57

\bibitem[Bennett et al.(2010)]{ob06109b}Bennett, D.P., Rhie, S.H., Nikolaev, S. et al. 2010, \apj, 713, 837


\bibitem[Bensby et al.(2013)]{bensby13} Bensby, T. Yee, J.C., Feltzing, S.\ et al.\ 2013, \aap, 549A, 147

\bibitem[Bensby et al.(2017)]{bensby17} Bensby, T. Feltzing, S., Gould, A.\ et al.\ 2017, \aap, submitted, arXiv:1702.02971

\bibitem[Bessell \& Brett(1988)]{bb88} Bessell, M.S., \& Brett, J.M.\ 1988, \pasp, 100, 1134

\bibitem[Bond et al.(2004)]{ob03235} Bond, I.A., Udalski, A., Jaroszy\'nski, M. et al.\ 2004, \apj, 606, L155


\bibitem[Calchi Novati et al.(2015a)]{21event}  Calchi Novati, S., Gould, A., Udalski, A., et al., 2015a, \apj, 804, 20

\bibitem[Calchi Novati et al.(2015b)]{170event} Calchi Novati, S., Gould, A., Yee, J.C., et al. 2015b, \apj, 814, 92

\bibitem[Calchi Novati \& Scarpetta (2016)]{scn16} Calchi Novati, S. \& Scarpetta, G. 2016, \apj, 824, 109



\bibitem[Clanton \& Gaudi(2014a)]{clanton14}Clanton, C.D. \& Gaudi, B.S. 2014a, \apj, 791, 90

\bibitem[Clanton \& Gaudi(2014b)]{clanton14b}Clanton, C.D. \& Gaudi, B.S. 2014b, \apj, 791, 91

\bibitem[Clanton \& Gaudi(2016)]{clanton16}Clanton, C.D. \& Gaudi, B.S. 2016, \apj, 819, 125

\bibitem[Dominik et al.(2007)]{signalman}Dominik, M. et al. Rattenbury, N.J., Allan, A., et al.\ 2007, \mnras, 380, 792



\bibitem[Dong et al.(2006)]{ob04343} Dong, S., DePoy, D.L., Gaudi, B.S., et al. 2006, \apj, 642, 842

\bibitem[Dong et al.(2007)]{smc001} Dong, S., Udalski, A., et al. 2007, \apj, 664, 862

\bibitem[Dong et al.(2009)]{ob05071b} Dong, S., Gould, A., Udalski, A., et al. 2009, \apj, 695, 970

\bibitem[Dong et al.(2009)]{mb07400} Dong, S., Bond, I.A., Gould, A., et al. 2009, \apj, 695, 970

\bibitem[Dominik(1999)]{dominik99} Dominik, M. 1999, \aap, 349, 108




\bibitem[Gaudi et al.(2008)]{ob06109}Gaudi, B.S., Bennett, D.P., Udalski, A. et al.\ 2008, Science, 319, 927

\bibitem[Gould(1992)]{gould92} Gould, A. 1992, \apj, 392, 442

\bibitem[Gould(1994a)]{gould94a} Gould, A. 1994a, \apjl, 421, L71

\bibitem[Gould(1994b)]{gould94b} Gould, A. 1994b, \apjl, 421, L75



\bibitem[Gould(2000)]{gould00} Gould, A. 2000, \apj, 542, 785

\bibitem[Gould(2004)]{gould04} Gould, A. 2004, \apjl, 606, 319

\bibitem[Gould(2008)]{gould08} Gould, A.\ 2008, \apj, 681, 1593

\bibitem[Gould \& Horne(2013)]{gouldhorne} Gould, A. \& Horne, K. 2013, \apjl, 779, L28

\bibitem[Gould \& Loeb(1992)]{gouldloeb} Gould, A. \& Loeb, A. 1992, \apj, 396, 104

\bibitem[Gould \& Yee(2012)]{gould12} Gould, A. \& Yee, J.C. 2012, \apj, 755, L17



\bibitem[Gould et al.(2013)]{prop2013} Gould, A., Carey, S., \& Yee, J. 2013, 2013spitz.prop.10036

\bibitem[Gould et al.(2014)]{prop2014} Gould, A., Carey, S., \& Yee, J. 2014, 2014spitz.prop.11006

\bibitem[Gould et al.(2015a)]{prop2015a} Gould, A., Yee, J., \& Carey, S., 2015a, 2015spitz.prop.12013

\bibitem[Gould et al.(2015b)]{prop2015b} Gould, A., Yee, J., \& Carey, S., 2015b, 2015spitz.prop.12015

\bibitem[Gould et al.(2016)]{prop2016} Gould, A., Yee, J., \& Carey, S., 2016, 2015spitz.prop.13005

\bibitem[Gould et al.(2010)]{gould10} Gould, A., Dong, S., Gaudi, B.S.\ et al.\ 2010, \apj, 720, 1073


\bibitem[Grether \& Lineweaver(2006)]{grether06} Grether, D., \& Lineweaver, C.H., 2006, \apj, 640, 1051

\bibitem[Griest \& Safizadeh(1998)]{griest98} Griest, K.\ \& Safizadeh, N.\ 1998, \apj, 500, 37


\bibitem[Han(2006)]{han06} Han, C.  2006, \apj, 638, 1080

\bibitem[Han et al.(2016)]{ob150479} Han, C., Udalski, A., Gould, A. et al.  2016, \apj, 834, 82

\bibitem[Henderson et al.(2016)]{henderson16} Henderson, C.B., Poleski, R., Penny, M. et al. 2016 \pasp 128, 124401

\bibitem[Huang et al.(2015)]{huang15} Huang, C.X, Penev, K., Hartman, J.D., et al. 2015, \mnras, 454, 4159


\bibitem[Jung et al.(2013)]{jung13}Jung, Y.\ K., Han, C., Gould, A., \& Maoz, D. 2013 \apjl, 768, L7

\bibitem[Kayser et al.(1986)]{kayser86} Kayser, R., Refsdal, S., \& Stabell, R. 1986, \aap, 166, 36

\bibitem[Kervella et al.(2004)]{kervella04} Kervella, P., Th{\'e}venin, F., Di Folco, E., \& S{\'e}gransan, D.\ 2004, \aap, 426, 297

\bibitem[Kim et al.(2016)]{kmtnet} Kim, S.-L., Lee, C.-U., Park, B.-G., et al.  2016, JKAS, 49, 37

\bibitem[Marcy \& Butler(2000)]{marcy00} Marcy, G.W. \& Butler, R.P. 2000, \pasp, 112,137

\bibitem[Martin et al.(2016)]{martin16} Martin, R.G., Livio, M., \& Palaniswamy, D. 2016, \apj, 832, 122 

\bibitem[Mao \& Paczy\'nski(1991)]{mao91} Mao, S.\ \& Paczy\'nski, B.\ 1991, \apj, 374, 37

\bibitem[Muraki et al.(2011)]{mb09266} Muraki, Y., Han, C., Bennett, D.P.,
et al.\ 2011, \apj, 741, 22

\bibitem[Nataf et al.(2013)]{nataf13} Nataf, D.M., Gould, A., Fouqu\'e, P. et al. 2013, \apj, 769, 88

\bibitem[Paczy\'nski(1986)]{pac86} Paczy\'nski, B.\ 1986, \apj, 304, 1



\bibitem[Pejcha \& Heyrovsk\'y(2009)]{pejcha09} Pejcha, O., \& Heyrovsk\'y, D.\ 2009, \apj, 690, 1772

\bibitem[Penny et al.(2013)]{penny13} Penny, M.T., Kerins, E., Rattenbury, N, et al.\ 2013, \mnras, 434, 2



\bibitem[Ranc et al.(2015)]{mb07197} Ranc, C., Cassan, A., Albrow, M.~D., et al.\ 2015, \aap, 580, A125


\bibitem[Refsdal(1966)]{refsdal66} Refsdal, S. 1966, \mnras, 134, 315


\bibitem[Ryu et al.(2017)]{ob160693} Ryu, Y.-H., Udalski, A., Yee, J.C., et al. 2017, in prep




\bibitem[Schneider \& Weiss(1988)]{schneider88}Schneider, P., \& Weiss, A. 1988, \apj, 330, 1

\bibitem[Shin et al.(2011)]{ob05018}Shin, I.-G., Udalski, A., Han, C., et al.\ 2011, \apj, 735, 855

\bibitem[Shin et al.(2012)]{ob110417}Shin, I.-G., Han, C., Choi, J.-Y., et al.\ 2012, \apj, 755, 91




\bibitem[Skottfelt et al.(2015)]{emccd} Skottfelt, I., Bramich, D.M., Hundertmark, M. et al. 2015, \aap, 574A, 54


\bibitem[Shvartzvald et al.(2016)]{shvartz16} Shvartzvald, Y., Maoz, D., Udalski, A.\ et al.\ 2016, \mnras, 457, 4089


\bibitem[Shvartzvald et al.(2017a)]{ukirt17} Shvartzvald, Y., Bryden, G.,
Gould, A.\ et al.\ 2017a, \aj, 135, 61

\bibitem[Shvartzvald et al.(2017b)]{ob161195} Shvartzvald, Y., Yee, J.C.,
Calchi Novati, S.\ et al.\ 2017b, \apjl, 840, L3

\bibitem[Skowron et al.(2011)]{ob09020}Skowron, J., Udalski, A., Gould, A et al.\ 2011, \apj, 738, 87




\bibitem[Soares-Furtado et al.(2017)]{soares17} Soares-Furtado, M., Hartman, J.D.; Bakos, G.A., et al. 2017, \pasp, 129, 4501

\bibitem[Spergel et al.(2013)]{spergel13} Spergel, D.N., Gehrels, N., Breckinridge, J., et al. 2013, arXiv:1305.5422


\bibitem[Street et al.(2016)]{ob150966} Street, R., Udalski, A., Calchi Novati, S.\ et al.\ 2016, \apj, 829, 93.

\bibitem[Suzuki et al.(2016)]{suzuki16} Suzuki, D., Bennett, D.P., Sumi, T. et al.\ 2016, \apj, 833, 145


\bibitem[Szymanski et al.(2011)]{ogleiii2} Szymanski, M., Udalski, A., Soszy\'nski, I., et al. 2011, Acta Astron., 61, 835

\bibitem[Thompson(2013)]{thompson13} Thompson, T.A. 2013, \mnras, 431, 63

\bibitem[Udalski(2003)]{ews2} Udalski, A. 2003, Acta Astron., 53, 291

\bibitem[Udalski et al.(1994)]{ews1} Udalski, A.,Szymanski, M., Kaluzny, J., Kubiak, M., Mateo, M.,  Krzeminski, W., \& Paczy\'nski, B. 1994, Acta Astron., 44, 317

\bibitem[Udalski et al.(2005)]{ob05071} Udalski, A., Jaroszy\'nski, M., Paczy\'nski, B, et al. 2005, \apj, 628, L109

\bibitem[Udalski et al.(2008)]{ogleiii1} Udalski, A., Szymanski, M.K., Soszynski, I., \& Poleski, R. 2008, Acta Astron 58, 69

\bibitem[Udalski et al.(2015a)]{ob140124} Udalski, A., Yee, J.C., Gould, A., et al. 2015, \apj, 799, 237

\bibitem[Wambsganss(1997)]{wambs97}Wambsganss, J. 1997, \mnras, 284, 172

\bibitem[Wo\'zniak(2000)]{wozniak2000} Wo\'zniak, P.~R. 2000, Acta Astron., 50, 421


\bibitem[Udalski et al.(2015b)]{ogleiv} Udalski, A., Szyma\'nski, M.K. \& Szyma\'nski, G. 2015b, Acta Astronom., 65, 1

\bibitem[Yee et al.(2012)]{mb11293} Yee, J.C., Shvartzvald, Y., Gal-Yam, A.\ et al.\ 2012, \apj, 755, 102

\bibitem[Yee et al.(2015a)]{yee15a} Yee, J.C., Udalski, A., Calchi Novati, S., et al., 2015a, \apj, 802, 76

\bibitem[Yee et al.(2015b)]{yee15} Yee, J.C., Gould, A., Beichman, C., 2015b, \apj, 810, 155

\bibitem[Yee et al.(2016)]{yee16} Yee, J.C., Johnson, J.A., Skowron, J., et al. 2016, \apj, 821, 121

\bibitem[Yoo et al.(2004)]{ob03262} Yoo, J., DePoy, D.L., Gal-Yam, A.\ et al.\ 2004, \apj, 603, 139



\bibitem[Zhu et al.(2015)]{ob141050} Zhu, W., Udalski, A., Gould, A., et al. 2015, \apj, 805, 8


\bibitem[Zhu et al.(2017a)]{zhu17} Zhu, W., Udalski, A., Calchi Novati, S., et al.\ 2017a, \aj, 154, 210 

\bibitem[Zhu et al.(2017b)]{zhu17b} Zhu, W., Huang, C., Udalski, A., et al. 2017b, \pasp, 129, 104501 

\bibitem[Zhu et al.(2017c)]{zhu17c} Zhu, W., Udalski, A., Huang, C.~X., et al.\ 2017c, \apjl, 849, L31


\end{thebibliography}
\end{document}